\documentclass[aps,prb,twocolumn,superscriptaddress,citeautoscript]{revtex4-1}

\usepackage{amsmath,amssymb}
\usepackage{bm}
\usepackage{graphicx}
\usepackage[table,xcdraw]{xcolor}
\usepackage{epstopdf}
\usepackage{latexsym}
\usepackage{subfigure}
\usepackage{color}
\usepackage{dsfont} %for ensemble alphabet, and works with numbers too, use \mathds{1}
\usepackage{wasysym} %for hexagon (in Eq environment or in text)
\usepackage{multirow}

\newcommand{\nccf}{{NaCaCo$_2$F$_7$}}
\newcommand{\nscf}{{NaSrCo$_2$F$_7$}}

\newcommand{\thickhline}{%
    \noalign {\ifnum 0=`}\fi \hrule height 1pt
    \futurelet \reserved@a \@xhline
}
\newcolumntype{"}{@{\hskip\tabcolsep\vrule width 1pt\hskip\tabcolsep}}
\makeatother

\begin{document}

\title{Single-ion properties of the $S_{\text{eff}}$ = 1/2 XY antiferromagnetic pyrochlores, \\ Na$A^{\prime}$Co$_2$F$_7$ ($A^{\prime} = $ Ca$^{2+}$, Sr$^{2+}$) }

\author{K.A. Ross}
%\affiliation{Colorado State University \textcolor{red}{make this current address}}
\affiliation{Colorado State University, Fort Collins, Colorado, 80523, USA}
\affiliation{Quantum Materials Program, Canadian Institute for Advanced Research, Toronto, Ontario M5G 1Z8, Canada}
\email{author to whom correspondence should be addressed:    \\kate.ross@colostate.edu}

\author{J.M. Brown} 
\affiliation{Colorado State University, Fort Collins, Colorado, 80523, USA}

\author{R.J. Cava} 
\affiliation{Princeton University, Princeton, New Jersey 08544, USA}
    
\author{J.W. Krizan} 
\affiliation{Princeton University, Princeton, New Jersey 08544, USA}

\author{S. E. Nagler} 
\affiliation{Quantum Condensed Matter Division, Oak Ridge National Laboratory, Oak Ridge, Tennessee 37831, USA}

\author{J.A. Rodriguez-Rivera} 
\affiliation{NIST Center for Neutron Research, National Institute of Standards and Technology, Gaithersburg, Maryland 20899, USA}
\affiliation{Materials Science and Engineering, University of Maryland, College Park, MD 20742, USA}

\author{M. B. Stone} 
\affiliation{Quantum Condensed Matter Division, Oak Ridge National Laboratory, Oak Ridge, Tennessee 37831, USA}

\date{\today}
\begin{abstract}

The antiferromagnetic pyrochlore material \nccf \ is a thermal spin liquid over a broad temperature range ($\approx$ 140 K down to $T_F  = 2.4 $K), in which magnetic correlations between Co$^{2+}$ dipole moments explore a continuous manifold of antiferromagnetic XY states \cite{ross2016static}.   The thermal spin liquid is interrupted by spin freezing at a temperature that is $\sim 2$ \% of the mean field interaction strength, leading to short range static XY clusters with distinctive relaxation dynamics.  Here we report the low energy inelastic neutron scattering response from the related compound \nscf, confirming that it hosts the same static and dynamic spin correlations as \nccf.   We then present the single-ion levels of Co$^{2+}$ in these materials as measured by inelastic neutron scattering.  An intermediate spin orbit coupling model applied to an ensemble of trigonally distorted octahedral crystal fields accounts for the observed transitions.  The single-ion ground state of Co$^{2+}$ is a Kramers doublet with a strongly XY-like $g$-tensor ($g_{xy}/g_{z} \sim 3$).  The local disorder inherent from the mixed pyrochlore $A$ sites (Na$^{+}$/Ca$^{2+}$ and Na$^{+}$/Sr$^{2+}$) is evident in these measurements as exaggerated broadening of some of the levels.  A simple model that reproduces the salient features of the single-ion spectrum produces approximately 8.4\% and 4.1\% variation in the $z$ and $xy$ components of the $g$-tensor, respectively.   This study confirms that an $S_{\text{eff}} =1/2$ model with XY antiferromagnetic exchange and weak exchange disorder serves as a basic starting point in understanding the low temperature magnetic behavior of these strongly frustrated magnets.  

\end{abstract}

%% thermal crossover from Heisenberg to XY??

\maketitle

\section{Introduction}
 \label{sec:intro}
\begin{figure}[!tb]  
\centering
\includegraphics[ width=\columnwidth]{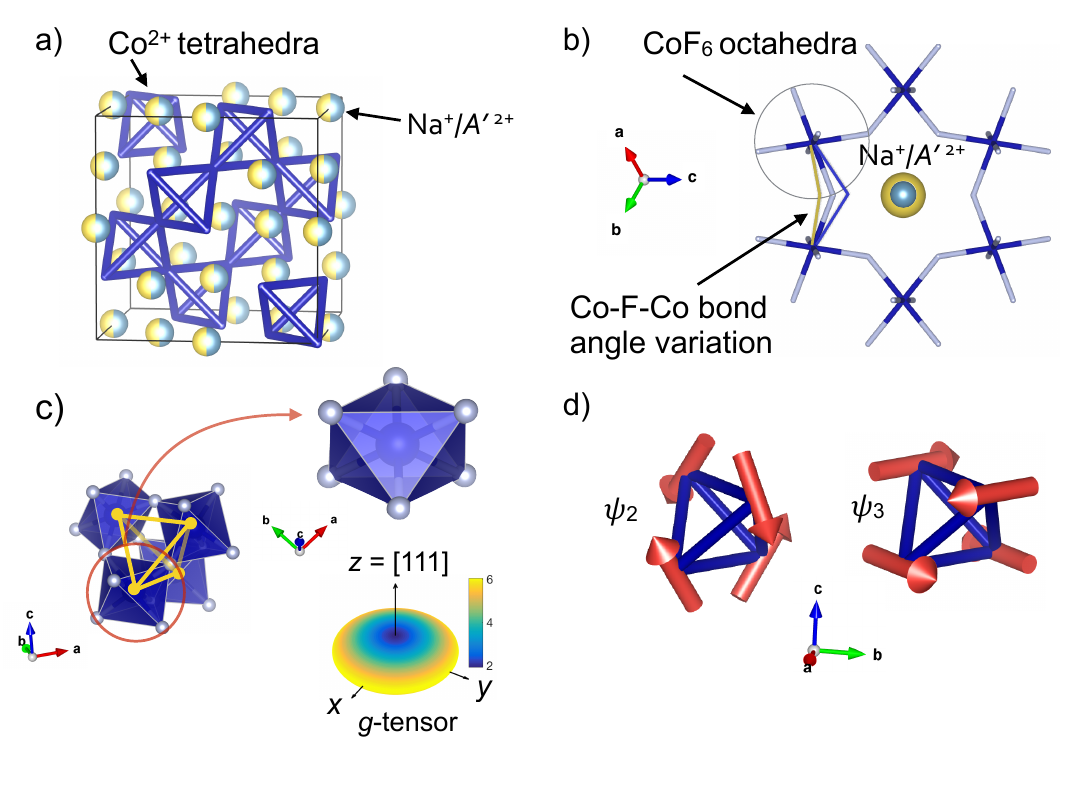}
\caption{a) Depiction of the two cation sublattices in \nccf \ and \nscf.  The $A  =  A^{\prime2+}$/Na$^+$ split site forms a pyrochlore sublattice (connectivity not shown here).  The nearest neighbor bonds in the $B = $Co$^{2+}$ pyrochlore sublattice are depicted.  b)  Relationship between the $A$ and $B$ sites and the average atomic positions.  The $A$ site cations reside in the open hexagons of the $B$ site pyrochlore structure, as viewed from the $\langle 111 \rangle$ body diagonals. The Co-F bonds are represented as rods.  The yellow and blue lines illustrate potential Co-F-Co bond angle variation with $A$ cation occupation at the central site.  c) Illustration of the average local CoF$_6$ octahedra, which are trigonally distorted along the local $\langle 111 \rangle$ directions that point into the center of a pyrochlore tetrahedron. The $g$-tensor is XY-like due to this environment.  d) The two basis states forming the $\Gamma_5$ manifold, from which the static and dynamic correlations are selected in \nscf \ and \nccf. }

\label{fig:struct}
\end{figure}

Magnetism on the pyrochlore lattice is a rich field of study that encompasses many unusual magnetic phenomena such as spin ice, spin liquids, and Order by Disorder (ObD) \cite{gardner2010magnetic}.    Experimental access to these phenomena has been granted mainly by rare earth oxide pyrochlore materials.  These have the general chemical formula $A_2B_2$O$_7$, with $A$ a magnetic trivalent rare earth cation and $B$ a non-magnetic tetravalent cation such as Ti, Sn, Zr, Ge or Pt  \cite{gardner2010magnetic, shannon1968synthesis, dun2015antiferromagnetic, cai2016high}.  Both the $A$ and $B$ sites independently form a pyrochlore-type sublattice, a highly-frustrated three dimensional network composed of corner-sharing tetrahedra (Figure \ref{fig:struct} a).   The anisotropic nature of the rare earth magnetic moments, arising due to strong spin-orbit coupling (SOC) combined with crystal electric field (CEF) effects, has offered fascinating variations  of magnetic behavior.  For instance, local Ising anisotropy is required for generating spin ice and its emergent magnetic monopoles in Ho$_2$Ti$_2$O$_7$ and Dy$_2$Ti$_2$O$_7$ \cite{castelnovo2012spin}, while XY anisotropy can lead to ObD \cite{bramwell1994order, wong2013ground, mcclarty2014order}, as suggested for the pyrochlore material Er$_2$Ti$_2$O$_7$ \cite{champion2003er, champion2003er, zhitomirsky2012quantum, savary2012order, oitmaa2013phase, maryasin2014order, andreanov2015order}.  However, the $4f$ electrons which are responsible for this magnetism have weak interaction strengths on the order of 1 K,  requiring experiments to be done at millikelvin temperatures in order to access their magnetic correlations.  This can limit the exploration of the lowest temperature states in these frustrated systems.  Furthermore, the ground states selected by these interactions can be quite sensitive to small amounts of chemical disorder, as has been observed in Yb$_2$Ti$_2$O$_7$ \cite{ross2012lightly, yaouanc2011single} and Tb$_2$Ti$_2$O$_7$ \cite{taniguchi2013long}. 

The recently discovered $3d$ transition metal pyrochlore fluoride \nccf \ (NCCF) has previously been suggested as a strongly interacting version of the antiferromagnetic (AFM) XY pyrochlore model \cite{ross2016static}, potentially allowing for a detailed investigation of this model and its variations at lower effective temperatures.    The XY anisotropy of effective spin 1/2 ($S_{\text{eff}} = 1/2$) magnetic moments in NCCF and the related compound \nscf \ (NSCF) are confirmed here through the modeling of Co$^{2+}$ single-ion levels by fitting to inelastic neutron scattering measurements.  Furthermore, both materials are shown to have nearly identical spin correlations in their spin frozen states, consisting of short range ordered XY clusters and associated dynamics, despite differences in the strength of chemical disorder.

NCCF and NSCF belong to a family of $3d$ transition metal fluoride pyrochlores \cite{krizan2014nacaco, krizan2015nasrco2f7, krizan2015nacan, sanders2016nasrmn2f7}.  The chemical formula and structure are similar to the rare earth oxide pyrochlores, taking the form $A_2B_2$F$_7$, where $A$ is a split site of Na$^+$/$A^{\prime}$  ($A^\prime$ = Ca$^{2+}$, Sr$^{2+}$) and $B$ is six-fold coordinated Co$^{2+}$, Ni$^{2+}$, Fe$^{2+}$ or Mn$^{2+}$ (Figure \ref{fig:struct}). Unlike the previously studied fluoride-pyrochlore CsNiCrF$_6$ \cite{zinkin1997short}, the magnetic $B$-site is chemically uniform, hosting just one magnetic species.  The average structures of these compounds as measured by x-ray diffraction show well-ordered pyrochlores (space group $Fd\bar{3}m$), consistent with random distribution of cations on the $A$ site.   Despite the materials being well-ordered on average, the local structure near the $B$ cations will be distorted away from the average trigonal $D_{3d}$ point group due to the $A$ site disorder, leading to variations in exchange parameters and single-ion anisotropy (Figure \ref{fig:struct} b).   In light of this, perhaps it is not surprising that all of the materials in this series eventually display spin freezing transitions.  However, the spin freezing occurs at very low \emph{effective} temperatures; their Curie-Weiss temperatures range from $\theta_{CW} \sim$ -70 \cite{sanders2016nasrmn2f7} to -140 K \cite{krizan2014nacaco}, but the spins freeze only at 2 to 4 K.  This gives large frustration indices of $f = \frac{|\theta_{CW}|}{T_F} \approx 19$ to 58, suggesting that the strength of the disorder, the presumed cause of the spin freezing, is weak compared to the overall interaction strength.   An interesting comparison in this regard can be made to the $B$-site disordered Yb-based pyrochlore Yb$_2$GaSbO$_7$; with its $\theta_{CW}= -1.15$ K and lack of a freezing transition down to the lowest measured temperature of 20 mK, this material demonstrates similarly ``ineffective'' exchange disorder relative to its (much weaker) average interaction strength ($f\ge 58$)\cite{blote1969heat, hodges2011magnetic}.

Unlike the rare earth oxides, nearly ideal Heisenberg moments with large spins should be expected for most of the members in this new series of transition metal fluoride pyrochlores.  This is due to the quenching of orbital angular momentum expected for most octahedrally coordinated $3d$ transition metals.  This means that phenomenology like spin ice or ObD will likely not pertain to most members of this series.   The exceptions to this are the Co$^{2+}$ compounds which we study here: NCCF and NSCF.   The free ion Co$^{2+}$ ($3d^7$) forms a $^4F$ ground term with $S=3/2$ and $L=3$, and when placed in a octahedral coordination the CEF and SOC conspire to form an $S_{\text{eff}}=1/2$ single-ion ground state \cite{buyers1971excitations, abragam2012electron}.  A distorted octahedral environment, such as the average environment of NCCF and NSCF, will lead to single-ion anisotropy in the $S_{\text{eff}} = 1/2$ states (Fig. \ref{fig:struct} c)\cite{lines1963magnetic, goff1995exchange, abragam2012electron}.  At temperatures low enough that the ground doublet states are the only relevant degrees of freedom, it is then possible model the interactions between effective spin 1/2 operators, in which the single-ion anisotropy encoded by the $g$-tensor is projected into the effective exchange interactions \cite{lines1963magnetic,white2006quantum,ross2011quantum,guitteny2013palmer}.  Thus, it seems that NCCF and NSCF could serve as new ``high temperature'' examples in which this successful method, well-developed for the rare earth oxide series, could be used \cite{ross2011quantum, savary2012order, yan2016general}.

 The static and dynamic spin correlations in NCCF were previously measured by inelastic neutron scattering \cite{ross2016static}.  It was found that below the freezing temperature ($T_F = 2.4$ K), short range order (SRO) of the spins develops with a correlation length of 16 \AA, corresponding to XY AFM configurations from the $\Gamma_5$ irreducible representation of the tetrahedral point group.  This manifold is spanned by two basis states called $\psi_2$ and $\psi_3$, shown in Figure \ref{fig:struct} d).  In the XY AFM pyrochlore model, the $q=0$ long range ordered (LRO) magnetic structures based on these two states are accidentally degenerate \cite{bramwell1994order, champion2004soft, wong2013ground, mcclarty2014order}, and either one can be selected by various ObD mechanisms, as has been discussed at length in the context of Er$_2$Ti$_2$O$_7$.   The observed selection of the non-coplanar $\psi_2$ state in that material \cite{poole2007magnetic} has been argued to occur either via quantum and thermal fluctuations (ObD) \cite{champion2003er, zhitomirsky2012quantum,savary2012order,oitmaa2013phase}, or from a non-ObD mechanism involving virtual excitations to higher crystal field levels \cite{petit2014order}.  \emph{Quenched} disorder in the form of dilution \cite{andreanov2015order} or exchange disorder \cite{maryasin2014order} has recently been predicted to select the coplanar $\psi_3$ state instead, and this has been studied in yttrium-diluted Er$_2$Ti$_2$O$_7$ \cite{gaudet2016magnetic}.  In NCCF, despite a clear mechanism for exchange disorder and the presence of XY AFM correlations, an LRO state is not selected.  
 
 %Presently it is unclear whether the SRO is purely $\psi_2$, $\psi_3$, or a distribution of linear combinations of these states.  Classical Monte Carlo simulations based upon a minimal XY AFM model suggest the latter interpretation \cite{sarkar2016unconventional}, but zero-field unpolarized neutron scattering does not distinguish between these possibilities.  
 
 The low energy inelastic neutron scattering (INS) response of NCCF shows non-dispersive, diffusive excitations with a distinctive intensity vs. $\vec{Q}$ pattern, extending to approximately $E =$ 10 meV \cite{ross2016static}, as well as a broad distribution of relaxation times in the $\mu$eV range, as probed by NMR \cite{sarkar2016unconventional}.  These dynamic signatures persist above the freezing temperature up to at least  $T=$14 K (INS \cite{ross2016static}) and 20 K (NMR \cite{sarkar2016unconventional}).  These data, combined with the high Curie-Weiss temperature ($\theta_{CW}$ = -140 K) relative to the freezing temperature ($T_F = 2.4$ K) giving a frustration index of $f = 58$, show that NCCF hosts an XY thermal spin liquid, i.e., a strongly correlated but disordered state dominated by entropy, over a large temperature range.  Surprisingly, NCCF resists ordering or freezing to much lower effective temperatures than the canonical XY pyrochlore Er$_2$Ti$_2$O$_7$ ($f = 20$)\cite{ruff2008spin}, suggesting that this material could be closer to a classical phase boundary in the general anisotropic exchange model developed for the rare earth oxides \cite{yan2016general}.  To make further progress in modeling NCCF, it is crucial to establish the relevance of the general anisotropic pseudo-spin 1/2 model.  This relies upon the understanding of the single-ion Hamiltonian for Co$^{2+}$ in these materials.

In this article we first present low energy inelastic and elastic neutron scattering measurements on the compound NSCF and compare it to the previously reported measurements on NCCF, demonstrating that despite their different $A^{\prime}$ sites the two compounds share the same experimental signatures and could be treated by the same theoretical approach.   We then present the observed single-ion levels for both compounds, measured by INS over the energy range of $E$ = 30 - 1500 meV. The observed INS response as well as the dc magnetic susceptibility is well described by a disorder-averaged intermediate SOC model.  This model confirms that the single-ion ground state in both materials is a well-isolated Kramers doublet, i.e., $S_{\text{eff}} = 1/2$, and provides the average $g$-tensors, which are observed to be strongly XY-like and have approximately 8\% variation due to local disorder.  

The contents of the paper are as follows: in Section \ref{sec:experimental} we give the experimental details of the INS measurements.  In Section \ref{sec:single-ion} we describe single-ion model used to fit the high energy INS data.  Section \ref{sec:lowEresults} demonstrates the equivalence of the spin correlations in the two materials, and presents detailed low energy INS measurements at temperatures above $T_F$ for the first time (i.e., the thermal spin liquid regime).  Section \ref{sec:singleionresults} details the single-ion results and fits.  The results are further discussed in Section \ref{sec:discussion}, and conclusions from this study are presented in Section \ref{sec:conclusions}.

%% this stuff can go later.

 \section{Experimental Method}
 \label{sec:experimental}
 
Single crystals of NCCF and NSCF (space group $Fd\bar{3}m$, room temperature lattice constants $a = 10.4189$ \AA \ and 10.545 \AA, respectively) were grown via the Bridgman-Stockbarger method in an Optical Floating Zone furnace, as previously reported \cite{krizan2014nacaco}.  

We studied a 3.527 g single crystal of NSCF using the MACS spectrometer at the NIST Center for Neutron Research \cite{rodriguez2008macs}.  The dynamic structure factor, $S(\vec{Q},E)$, was measured in the [$HHL$] reciprocal lattice plane.  Neutrons with a final energy $E_f = 3.7$ meV were selected, and post-sample BeO filters were used to remove higher harmonic contamination and reject neutrons for which $E_f>3.7$ meV.  For elastic scattering, a Be filter preceded the sample, while for inelastic scattering no incident filters were used.   The resulting energy resolution was $\delta E$ = 0.17 meV at the elastic line.  These data are compared to those previously published on a 0.87 g crystal of NCCF, which was also taken using the MACS spectrometer in a similar configuration \cite{ross2016static}.  The relative masses cannot be used to compare the intensities directly due to differences in crystal mounting.  The NSCF and NCCF data were scaled relative to each other to match the inelastic intensities near (002) and (111) (Fig. \ref{fig:MACS}).  The same scaling also produced matching (220) magnetic intensities below $T =$ 1.7 K.  

In order to investigate the Co$^{2+}$ single ion levels in NCCF and NSCF, higher energy INS experiments were performed using the SEQUOIA time-of-flight chopper spectrometer at the Spallation Neutron Source, Oak Ridge National Laboratory \cite{granroth2006sequoia}.   The crystals were oriented with [HHL] in the horizontal scattering plane.  Data were taken at $T$ = 5 K and 200 K, using incident energies of $E_i =$ 60, 250, 700, and 2500 meV.  This wide range of incident energies was employed to probe the large dynamic range expected for single ion energy levels of Co$^{2+}$. For these energies, the T$_0$ Chopper and Fermi Chopper (FC1 or FC2) speeds were set to the following values: for $E_i$ = 60 meV, 60 Hz (T$_0$) and 420 Hz (FC2); for $E_i$ = 250 meV,  120 Hz (T$_0$) and 480 Hz (FC1); for $E_i$ = 700 meV, 150 Hz (T$_0$) and 600 Hz (FC1); and for $E_i$ = 2500 meV, 180 Hz (T$_0$) and 600 Hz (FC1).  The elastic energy resolutions in these configurations were $\delta E$ = 1.92, 12.71, 44.96 and 276.44 meV, respectively.  Note that $\delta E$ decreases as the neutron energy transfer increases for each $E_i$, and the full energy-dependent resolution for SEQUOIA \cite{granroth2006sequoia} was employed in our fits.

\section{Single-ion calculations for Co$^{2+}$}
\label{sec:single-ion}

In this section we present a method for calculating the single-ion levels in Co$^{2+}$ within an intermediate spin orbit coupling scheme, as well as the resulting INS response and dc magnetic susceptibility. This method was used to fit the data presented in Section \ref{sec:singleionresults}.  The method is similar to that used for the rare earth based magnets \cite{gaudet2015neutron, babkevich2015neutron}, and results in similar phenomenology.  The single ion levels for Co$^{2+}$ in a trigonally distorted octahedron are Kramers doublets with anisotropic moments described by a $g$-tensor.

%If, in the temperature range of interest, the vast majority of Co$^{2+}$ ions are in the ground state doublet (i.e. $k_B T \ll E_1$, the energy of the next excited single-ion level), the Co$^{2+}$ moments can be treated as $S_\text{eff}$ = 1/2.  The moment size will be anisotropic for point group symmetry lower than $O_h$, as encoded by the $g$-tensor, which can be calculated from the single-ion ground state doublet, as described below.

%This means that, as long as the requirement for a well-isolated CEF ground state is satisfied, the average interactions in NCCF and NSCF may also be treated with the anisotropic exchange $S_\text{eff}$ =  1/2 Hamiltonian that has been used to successfully describe various magnetic phenomena in rare earth pyrochlores \REF[all the refs]

\subsection{Intermediate spin orbit coupling Hamiltonian}
 \label{sec:hamiltonian}
The single-ion Hamiltonian is approximated by two contributions,

\begin{equation}
 H_\text{ion} = H_\text{CEF}(\vec{L}) + H_{\text{SOC}}(\vec{L},\vec{S}),
 \label{eqn:Hion}
 \end{equation}

  where $H_\text{CEF}(\vec{L})$ is the crystal electric field (CEF) Hamiltonian that acts only on the orbital angular momentum subspace, and H$_\text{SOC}$ is the spin orbit coupling (SOC) term. The CEF Hamiltonian can be written in general as:

\begin{equation}
 H_{\text{CEF}} =  \sum_{l,m} B_{l,m} \hat{O}_{l,m} 
 \label{eqn:Hcef}
 \end{equation}

where $\hat{O}_{l,m}$ are the Stevens operator equivalents \cite{stevens1952matrix, hutchings1964point, mcphase_manual}.  For transition metal ions, these operators are written in terms of the orbital angular momentum matrix operators $\hat{L_+}$, $\hat{L_-}$ and $\hat{L_z}$ (as opposed to the total angular momentum operators  $\hat{J_+}$, $\hat{J_-}$ and $\hat{J_z}$ relevant for $f$-electron systems)\cite{hutchings1964point}.   For the trigonal point group symmetry relevant to the average local environments of Co$^{2+}$ in NCCF and NSCF, $(l,m)= (2,0),(4,0)$ and $(4,3)$ are the only nonzero terms \cite{hutchings1964point}.  The $B_{l,m}$ values can be extracted by fitting to INS data.
%, and the results are described in Section \ref{sec:singleionresults}.

The spin orbit coupling term is given by,

\begin{equation}
 H_{\text{SOC}} = p \lambda (\vec{S}\cdot \vec{L}) = p \lambda(\hat{S}_x \hat{L}_x + \hat{S}_y \hat{L}_y + \hat{S}_z \hat{L}_z),
 \label{eqn:Hsoc}
 \end{equation}

where for the free ion Co$^{2+}$ $\lambda = -22.32$ meV \cite{kanamori1957theory}, and $p$ is the ``orbital reduction parameter'' that can be used to account for changes in effective SOC strength due to covalency \cite{buyers1971excitations}.  The $x$ and $y$ spin operators are linear combinations of $\hat{S}_+$ and $\hat{S}_-$ as usual:  $\hat{S}_x = \frac{1}{2}(\hat{S}_+ + \hat{S}_-)$ and $\hat{S}_y = \frac{1}{2i}(\hat{S}_+ - \hat{S}_-)$ and similarly for $\hat{L}_x$ and $\hat{L}_y$. \\
\\ 

The full single-ion Hamiltonian (Eqn. \ref{eqn:Hion}) can be diagonalized within the $28 \times 28$ manifold of states formed by the $|L_z, S_z\rangle$ basis of the $S=3/2$ and $L=3$ free ion term ($^4F$) for Co$^{2+}$.  A trigonally distorted octahedral coordination, as found in NCCF and NSCF, produces 14 Kramers doublets.  The presence of a doublet ground state permits the description of the magnetic moments as pseudo-spin 1/2 ($S_{\text{eff}} = 1/2$) at sufficiently low temperatures.  This doublet can be described as an anisotropic magnetic moment with strength given by the $g$-tensor \cite{abragam2012electron}.  In the trigonal symmetry appropriate to NCCF and NSCF, the $g$-tensor has two independent components, $g_z$ and $g_{xy}$, which refer, respectively, to the local [111] direction pointing into the center of the tetrahedron, and the plane perpendicular to that (Fig. \ref{fig:struct}c).   The $g$-tensor components are given by the matrix elements of the magnetic moment operator in the subspace of the two ground state wavevectors ($|\nu_1 \rangle$ and $|\nu_2\rangle$),
 
\[g_{xy} = -2 \langle\nu_1| (\hat{L}_x + 2 \hat{S}_x)|\nu_2\rangle,\] and
\[g_{z} = 2 \langle\nu_2| (\hat{L}_z + 2 \hat{S}_z)|\nu_2\rangle. \]

The resulting saturated magnetic moment sizes in the principal directions are $\mu_z = \frac{g_z}{2} \mu_B$ and $\mu_{xy} = \frac{g_{xy}}{2} \mu_B$.

\begin{figure}[!tb]  
\centering
\includegraphics[ width=\columnwidth]{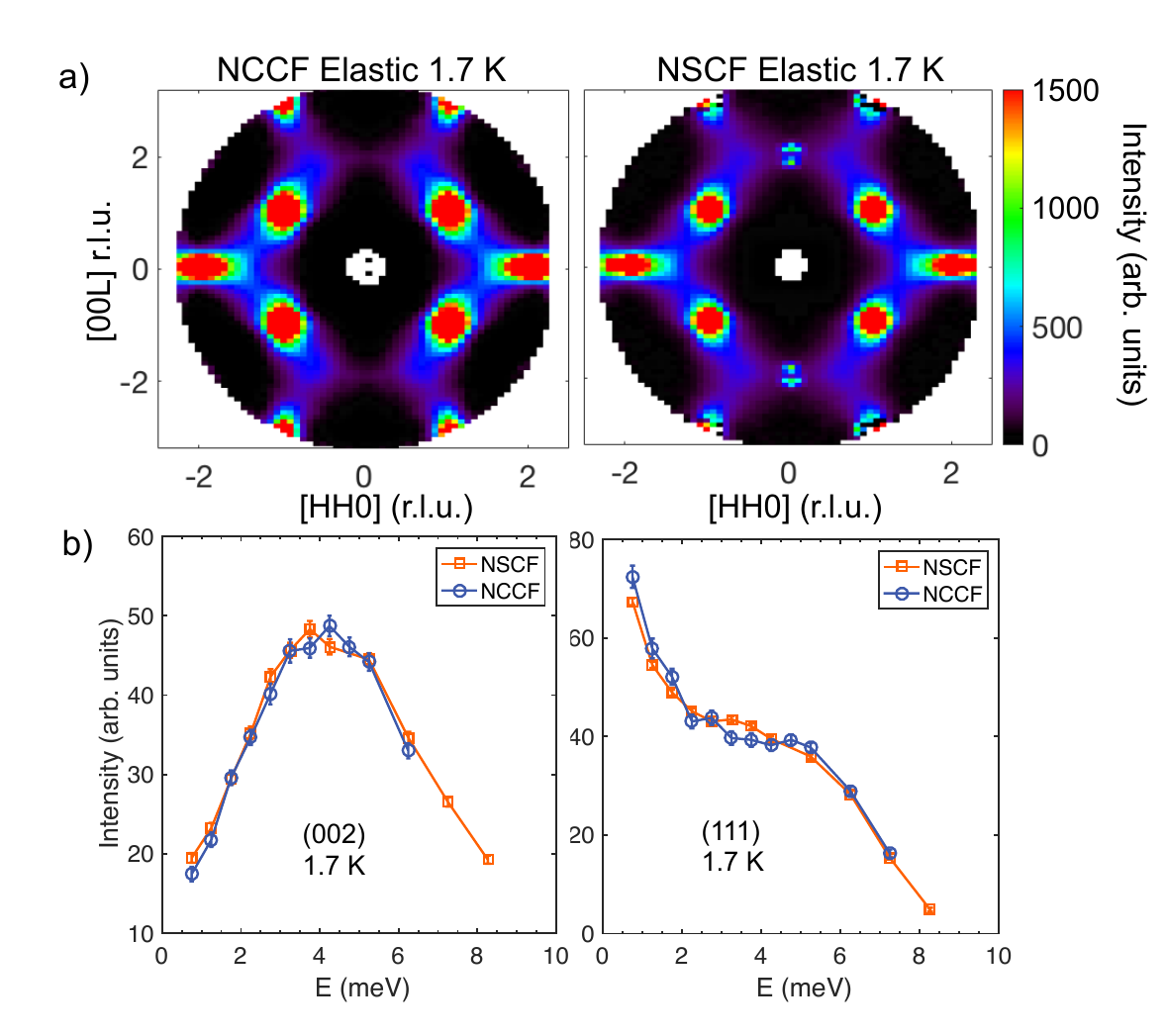}
\caption{ Comparison of neutron scattering intensities from NSCF and NCCF.  Data taken on the MACS spectrometer with energy resolution $\delta E=0.17$ meV.  a) Magnetic elastic scattering in \nccf \ (left) and \nscf \ (right) at $T$ = 1.7 K.  Intensities from symmetrically equivalent quadrants of the scattering plane were averaged.  The sharp feature near (002) in NSCF is due to multiple scattering and is not representative of spin correlations.  b) Constant-$\vec{Q}$ scans near the (002) and (111) positions.  Errorbars represent one standard deviation.}

\label{fig:MACS}
\end{figure}

\subsection{Comparison to inelastic neutron scattering} 
 \label{sec:INS_comparison}
The dynamic structure factor, $S(\vec{Q},E)$, is related to the observed intensity of inelastic neutron scattering via $I(\vec{Q},E) = \frac{k_f}{k_i} f(|Q|)^2 S(\vec{Q},E)$, where $k_i$ and $k_f$ are the initial and final wavenumbers of neutrons scattered with energy transfer $E = \frac{\hbar}{2m} (k_i^2 - k_f^2)$, and $f(|Q|)$ is the magnetic form factor \cite{squires2012introduction}.  The single-ion dynamic structure factor at constant $|Q|$ associated with Eqns. \ref{eqn:Hion} to \ref{eqn:Hsoc} can be calculated as follows \cite{jensen1991rare}:

\begin{equation}
 S(E) = C \sum_{n,n',\alpha}\frac{e^{-\beta E_n}}{Z}\frac{\Gamma_{\langle n,n' \rangle} |\langle \nu_n | \hat{L}_\alpha + 2 \hat{S}_\alpha | \nu_{n'} \rangle|^2}{([E_{n'} - E_{n}] - E)^2 + \Gamma_{\langle n,n' \rangle}^2},
 \label{eqn:SofE}
 \end{equation}

where $\alpha = x,y,z$, while $n$ and $n'$ label eigenstates of Eqn. \ref{eqn:Hion}.  $C$ is a scale factor applied to match the measured intensity of constant $|Q|$ data (note that at constant $|Q|$ the form factor $f(|Q|)^2$ can be absorbed into $C$).  The partition function is $Z = \sum_n \exp{(-\beta E_n)}$ with $\beta = 1/k_B T$. $|\nu_n\rangle$ is the wavevector from the single ion calculation corresponding to energy eigenvalue $E_n$.  A Lorentzian half width at half maximum (HWHM), $\Gamma_{\langle n,n' \rangle}$, accounts for line broadening due to instrumental resolution, finite excitation lifetimes, or averaged dispersion of the transitions between state $n$ and $n'$.  To fit the data presented in Section \ref{sec:singleionresults}, we set $\Gamma_{\langle n,n' \rangle}$ to the energy transfer-dependent instrument resolution for all transitions except those involving the ground doublet, $n = 1,2$, to the first excited doublet $n' = 3,4$.  In that case, $\Gamma_{\langle n,n' \rangle} = 13.5$ meV accounts for the increased width due to the finite dispersion of the first excited level (Fig. \ref{fig:firstlevel}).   An elastic peak with a full width at half maximum (FWHM) fixed at the instrumental resolution was also added to account for all sources of nuclear elastic scattering, both coherent and incoherent.

 \begin{figure*}[!tb]  
\centering
\includegraphics[ width=2\columnwidth]{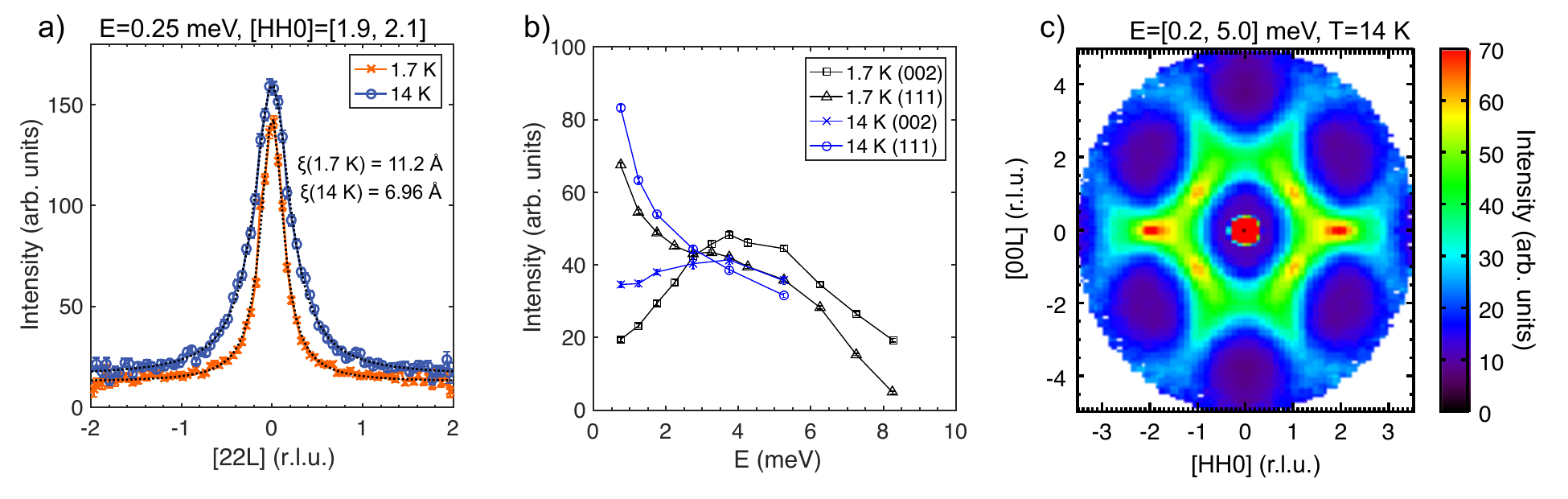}
\caption{Temperature dependence of inelastic scattering in NSCF. a) Constant energy cuts at 0.25 meV along [22L] reveal how the correlation length of the low energy dynamics changes with increasing temperature.  The dynamic correlation length along this direction decreases from 11.2 \AA \ to 6.96 \AA \ upon heating (dotted lines are fitted Lorentzians).  b) Constant $Q$ cuts near the (002) and (111) Bragg positions at T=14K and 1.7K.  c)  Diffuse inelastic scattering in NSCF above the freezing temperature, at $T$=14 K (data folded and symmetrized, integrated from $E$ = 0.2 to 5.0 meV).  Errorbars represent one standard deviation.}
\label{fig:Tdep}
\end{figure*}

To account for disorder in the local environment of Co$^{2+}$ brought about by the randomly mixed $A$ site in the crystal structures of NCCF and NSCF, the average $S(E)$ can be calculated over an ensemble of ions with slightly varying $B_{l,m}$ parameters in Eqn. \ref{eqn:Hcef}.  If the local site symmetry remains trigonal, as expected based on results of vibrational spectroscopy from the related compound NaCaMg$_2$F$_7$ \cite{oliveira2004crystal}, there remain only three parameters that enter into $H_{\text{CEF}}$.  For simplicity we chose a linear distribution of the $B_{4,0}$ parameter, allowing the width of the distribution ($\Delta B_{4,0}$) to be fit.  The reasoning for this choice is described in Section \ref{sec:discussion}. 

\subsection{Magnetic Susceptibility} 
 \label{sec:suscept}

The eigenvectors and eigenvalues of Eqn. \ref{eqn:Hion} also permit the calculation of the (powder averaged) single ion magnetic susceptibility as a function of temperature \cite{jensen1991rare, kimura2014experimental},

\begin{multline}
 \chi_{\text{ion}}(T) = \frac{N_A \mu_B^2}{3 k_B Z} \sum_\alpha \bigg(\frac{ \sum_{n} |\langle \nu_n | \hat{L}_\alpha + 2\hat{S}_\alpha | \nu_n \rangle|^2 e^{-E_n/T}}{T}  \\ + \sum_n\sum_{m\neq n}  |\langle \nu_m | \hat{L}_\alpha + 2\hat{S}_\alpha | \nu_n \rangle|^2 \frac{e^{-E_n/T} - e^{-E_m/T}}{E_m - E_n} \bigg). 
\label{eqn:suscept}
\end{multline}

  This expression includes the van Vleck susceptibility, but excludes the diamagnetic susceptibility $\chi_{\text{dia}}$, and does not account for the mean field interaction between magnetic moments.  To compare to the measured dc susceptibility, the following mean field approximation can be used \cite{white2006quantum},

\begin{equation}
\chi_{\text{MF}} = \chi_{\text{dia}} + \frac{\chi_{\text{ion}}}{1 + \lambda_W \chi_{\text{ion}}}.
\label{eqn:chiMF}
\end{equation}

Here, $\lambda_W$ is the Weiss molecular field constant, which accounts for the mean exchange interactions.  With this sign convention, a positive value of $\lambda_W$ indicates AFM interactions.

\section{Results}
\label{sec:results}
\subsection{Low energy dynamic structure factor in \nscf} 
 \label{sec:lowEresults}

The low temperature ($T = 1.7$ K) and low energy ($E < 5$ meV) magnetic neutron scattering response from NSCF is compared to that from NCCF in Figure \ref{fig:MACS}.  The NCCF data was previously published in Ref. \onlinecite{ross2016static}.   For the elastic scattering presented for both materials, the equivalent high temperature ($T = 14$ K $> T_F$) elastic maps were subtracted in order to isolate the magnetic scattering.   For the inelastic scattering, an empty can subtraction was performed.   The neutron scattering intensity of NCCF has been scaled such that the (220) magnetic elastic peak intensity matches NSCF, which also produces equivalent intensities for the inelastic scattering.  The neutron scattering patterns are nearly identical over the whole energy range, indicating that the static and dynamic spin correlations are equivalent in both compounds.  An interpretation of this scattering  in terms of static and dynamic XY correlations has been presented in Ref. \onlinecite{ross2016static}; this interpretation is expected to qualitatively apply to both materials.  The sharp features near the (002) and (00$\bar{2}$) positions in NSCF seen in Fig. \ref{fig:MACS} a) appear in the elastic channel using several choices of incident energies, but can be made to vanish for other choices, signifying that their origin is multiple scattering.   Such (002) scattering has been observed before in rare earth titanates and was previously thought to be caused by symmetry lowering, but has recently been identified as multiple scattering in those materials as well \cite{baroudi2015symmetry}.  

The data shown in Fig. \ref{fig:MACS} was collected below the freezing temperatures of NCCF and NSCF (2.4 K and 3.0 K, respectively).  Above the freezing transition, it was shown previously in NCCF that the elastic magnetic scattering vanishes, while inelastic magnetic scattering at E = 0.5 meV persists to at least $T=14$ K and retains the same distinctive pattern in reciprocal space \cite{ross2016static}.  In NSCF we have confirmed the persistence of this inelastic scattering above $T_F$ at $T=14$ K.  The inelastic scattering at $E=0.25$ meV is broader at 14 K compared to 1.7 K, implying shorter ranged dynamic correlations (Fig. \ref{fig:Tdep} a).  The full energy dependence of the scattering up to 5 meV is compared for the two temperatures in Fig. \ref{fig:Tdep} b), which shows that the spectrum has similar features above and below $T_F$, but with enhanced low energy spectral weight at $T=$14 K.  The energy integrated scattering at $T = 14$ K in NSCF ($E = 0.2$ to $5.0$ meV) over the [HHL] plane is shown in Fig. \ref{fig:Tdep} c).  This scattering could be compared to equal time correlation functions, such as those calculated in Ref. \onlinecite{yan2016general}.

\subsection{Single ion levels and average $g$-tensors of \nscf \ and \nccf}
 \label{sec:singleionresults}
 
\begin{figure}[!tb]  
\centering
\includegraphics[ width=\columnwidth]{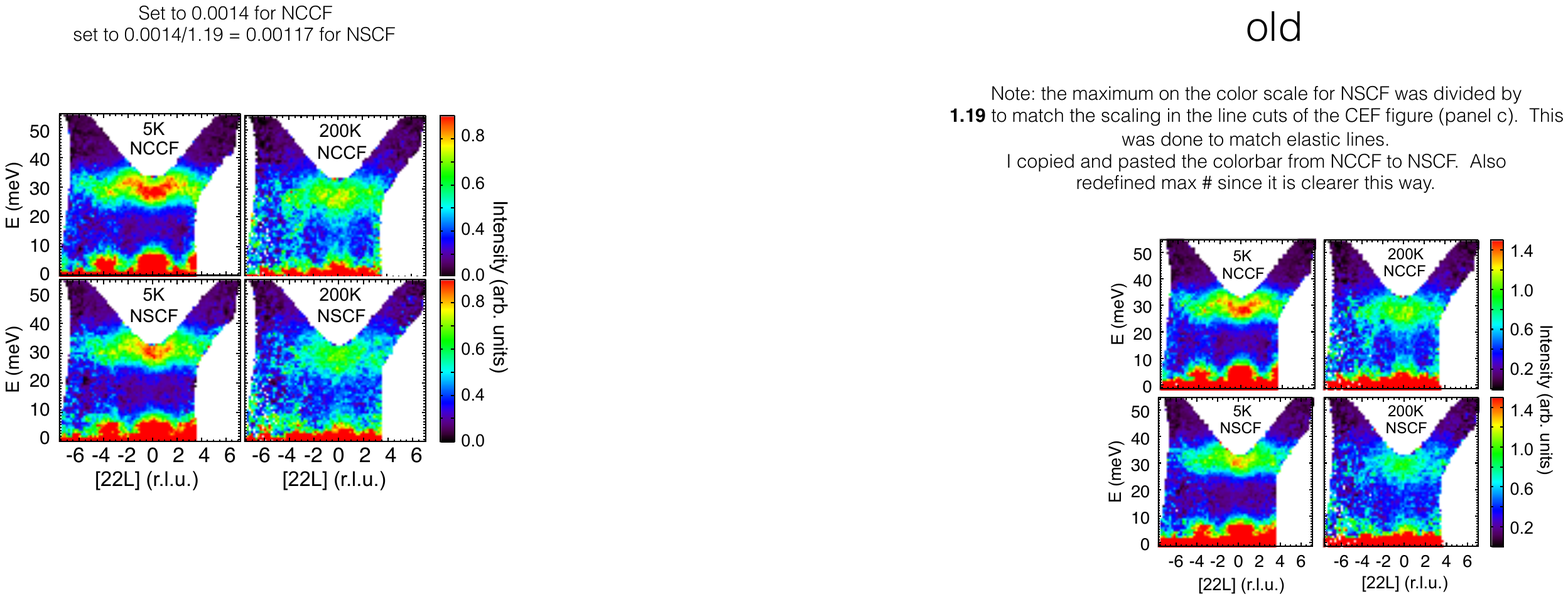}
\caption{The first excited single ion levels in NCCF (top row) and NSCF (bottom row), observed by inelastic neutron scattering ($E_i = 60$ meV). Data are averaged over the following ranges in perpendicular reciprocal lattice directions (r.l.u.): [HH0], H= [1.9, 2.1] and [K-K0], K= [-0.1, 0.1]. }
\label{fig:firstlevel}
\end{figure}

\begin{figure*}[!tb]  
\centering
\includegraphics[ width=2\columnwidth]{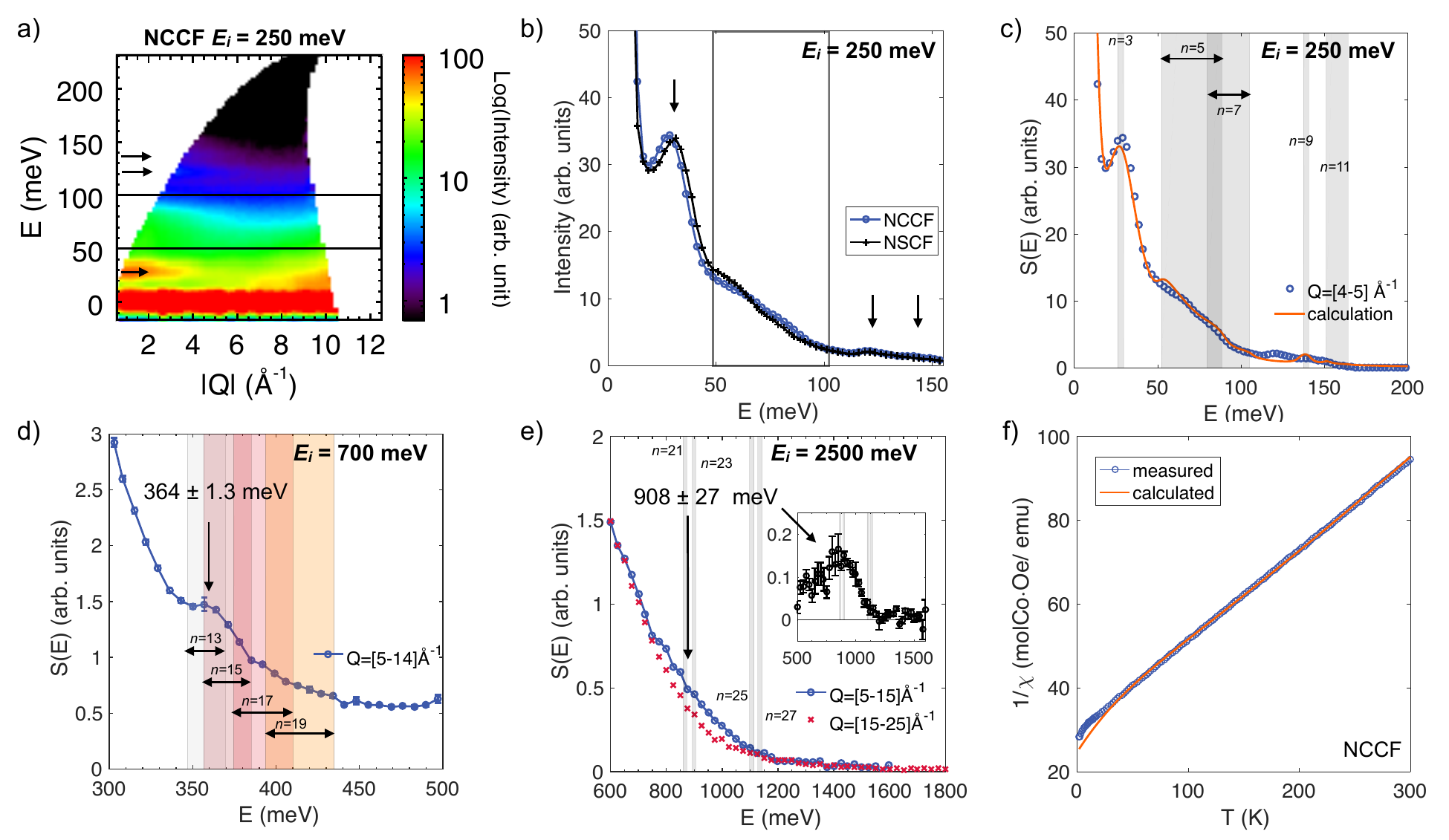}
\caption{  Measurements and fits of single ion levels of Co$^{2+}$ in \nccf.  Panels a) through e) show the observed single-ion levels from orientationally averaged single crystal inelastic neutron scattering measurements (SEQUOIA spectrometer), at $T$= 5 K. Shaded regions in panels c) to e) indicate the energy ranges spanned by the doubly-degenerate eigenvalues of Eqn. \ref{eqn:Hion}; the eigenvalues vary for different local environments, producing a range of values (see main text).  The first eigenstate label, $n$, in the corresponding doublet is indicated for each range.  a) Energy vs. $|Q|$ pseudo-color plot, with $E_i = 250$ meV.  Intensity is displayed on a logarithmic scale.  Arrows indicate the positions of relatively sharp magnetic modes, and the rectangle indicates a broad band of magnetic scattering. b) Cut through data in panel a), and equivalent data from NSCF, averaging from $|Q|$ = 4 - 5 \AA$^{-1}$ (empty can background subtracted).  c)  Fit to the NCCF data shown in panel b).  d)  $E_i$ = 700 meV, averaging from $|Q| = 5-14$ \AA$^{-1}$. e) $E_i$ = 2500 meV, cuts taken at both high and low $|Q|$.  On top of a steep non-magnetic background, additional intensity is seen for low $|Q|$ near 900 meV.  The inset shows the subtraction of the high-$|Q|$ background, which gives a peak at 908 $\pm$ 27 meV. f)  Single-ion susceptibility with mean field interactions  (Eqns. \ref{eqn:suscept} and \ref{eqn:chiMF}) compared to the measured dc susceptibility of NCCF from Ref. \onlinecite{krizan2014nacaco}. }

\label{fig:CEF}
\end{figure*}

Transitions between single ion levels of Co$^{2+}$ were observed in NCCF and NSCF using the SEQUOIA time of flight neutron spectrometer.   The energies of the observed transitions were found range from 28  to 908 meV, requiring the use of multiple incident energies.   The first excited level measured using $E_i$ = 60 meV in both NCCF and NSCF is shown in Figure \ref{fig:firstlevel}, at both $T = 5$ and 200 K.   In both materials this level is located near 30 meV, persists to temperatures above $|\theta_{CW}|$ (demonstrating its single-ion origin), and displays the characteristic decreasing intensity vs. $|Q|$ dependence of magnetic excitations (Fig. \ref{fig:CEF}a)).  At $T = 5$ K the first excited level displays distinct dispersion, suggesting exchange-induced mixing of the ground state and first excited level \cite{buyers1971excitations}.    Note that the low energy spin excitations presented in Section \ref{sec:lowEresults} are also identifiable in Fig. \ref{fig:firstlevel} below $\sim$ 10 meV; these are due to the magnetic correlations discussed in Ref. \onlinecite{ross2016static}, rather than single-ion levels.

The full set of observed single-ion levels in NCCF at $T=5$ K are shown in Figure \ref{fig:CEF} along with results of fitting the model described in Section \ref{sec:single-ion}.  Some of these features are significantly broader than the energy resolution as expected, since the local variation in Co environment should produce finite energy ranges over which transitions are observed.   The measured energies of the single-ion levels and corresponding energy widths are tabulated in Table \ref{tab:measured_peaks} for both NCCF and NSCF.\footnote{The  $E \sim 360$ meV mode was measured only in NCCF due to time limitations.  Given the overall similarity if the other single-ion levels in the two materials, this mode is assumed to lie in same energy range in NSCF.}   All neutron scattering data in Fig. \ref{fig:CEF} have a corresponding background subtracted and are corrected for the $k_f/k_i$ factor described in Section \ref{sec:INS_comparison}, thus representing the quantity $f(|Q|)^2 S(Q,E)$ in arbitrary units.   The orientational average of these single crystal data are presented in order to provide a concise overview of the excitations; no significant dispersion was detectable in any but the first excited level, and this dispersion is well-accounted for in the model by an increased FWHM, as discussed below.    For a uniform Co$^{2+}$ environment, 14 levels (including the ground state) are expected in total, each of them doubly degenerate according to Kramers theorem.  Note that at $T = $ 5 K, a low temperature compared to the energy of the first excited level ($\sim 300$ K), only transitions from the ground state to the excited states will be observable.   Additionally, some of these transitions may have weak intensities, depending on the strength of the transition matrix elements between eigenstates of Eqn.~\ref{eqn:Hion}. The intensities are weighted by the square of the transition matrix elements of the magnetic moment operator, $M^2_{\langle n,n' \rangle}  = \sum_\alpha |\langle \nu_n | \hat{L}_\alpha + 2 \hat{S}_\alpha | \nu_{n'} \rangle|^2$, which enter into Eqn.~\ref{eqn:SofE}.  
 
\begin{table}[]
\centering
\begin{tabular}{|l|l|l|l|}
\hline
\multicolumn{2}{|c|}{NCCF}  & \multicolumn{2}{c|}{NSCF}                           \\ \hline
Energy (meV) & FWHM (meV)   & Energy (meV)             & FWHM (meV)               \\ \hline
28.05(2)$^a$     &  27.50(7)           & 29.31(2)$^a$                 & 29.53(8)                       \\ \hline
$\sim$ 46 - 101$^a$     & broad        & $\sim$ 41 - 100$^a$                 & broad                    \\ \hline
121.9(1)$^a$    & 9.1(4) & 120.7(2)$^a$                 & 7.2(9)$^*$             \\ \hline
$\sim$ 139$^a$          & broad        & $\sim$ 134$^a$                      & broad                    \\ \hline
364(1)$^b$       & 24(3)           & -- & -- \\ \hline
9.1(3)$\times$10$^{2}$ $^c$     & 3.2(6)$\times$10$^{2}$          & 9.0(3)$\times$10$^{2}$  $^c$               & 3.7(6)$\times$10$^{2}$                    \\ \hline
\end{tabular}
\caption{Measured single-ion levels in \nccf \ (NCCF) and \nscf \ (NSCF), with central peak positions and full width at half maxima (FWHM). Due to time constraints, the peak expected near 360 meV in NSCF was not measured. The quoted error on the peak centers is the standard deviation of the fitted peak center.  The FWHM are marked with an asterisk ($^*$) when they are resolution-limited.  Instrument configurations are indicated by $a : E_i = 250$ meV, $b : E_i = 700$ meV, $c : E_i = 2500$ meV. }
\label{tab:measured_peaks}
\end{table}

   Fig. \ref{fig:CEF} a) shows measurements at $E_i = 250$ meV, with the intensity presented on a logarithmic color scale.  This shows several of the magnetic excitations, including the first excited level near 30 meV (as in Fig. \ref{fig:firstlevel}), a broad band of magnetic features spanning approximately 46 to 100 meV (indicated by a rectangle), and two higher energy features near 120 and 140 meV.   Note that phonon scattering is also visible below 30 meV, but its intensity increases as function of $|Q|$, while the  magnetic features show the opposite trend in intensity vs. $|Q|$ due to the magnetic form factor (see also Appendix \ref{sec:Qdep}).  Intensity vs. energy cuts from $Q=[4,5]$ \AA$^{-1}$ are compared for NCCF and NSCF in Fig. \ref{fig:CEF} b), showing the overall similarity between the single ion levels in the two compounds.  The first excited level is at a slightly higher energy for NSCF (29.31(2) meV) compared to NCCF (28.05(2) meV).   Two higher energy levels are shown in Fig. \ref{fig:CEF} d) and e), located at $364 \pm 1$ meV and $908 \pm 30$ meV.  In these high energy ranges, multi-phonon scattering creates a strong, sloping background and obscures these magnetic features.  Nevertheless, their magnetic nature is confirmed by their $|Q|$-dependence.  For the data shown in panel d), the single-ion level energy was extracted using a fit that included a sloping background plus a Gaussian.   For the 908 meV feature, intensity vs. $E$ for two $|Q|$ ranges were compared, as shown in panel e).   The precision of the energy determination for these higher modes is relatively poor due to decreased energy resolution of the higher incident energy instrument configurations.

\begin{figure}[!tb]  
\centering
\includegraphics[ width=\columnwidth]{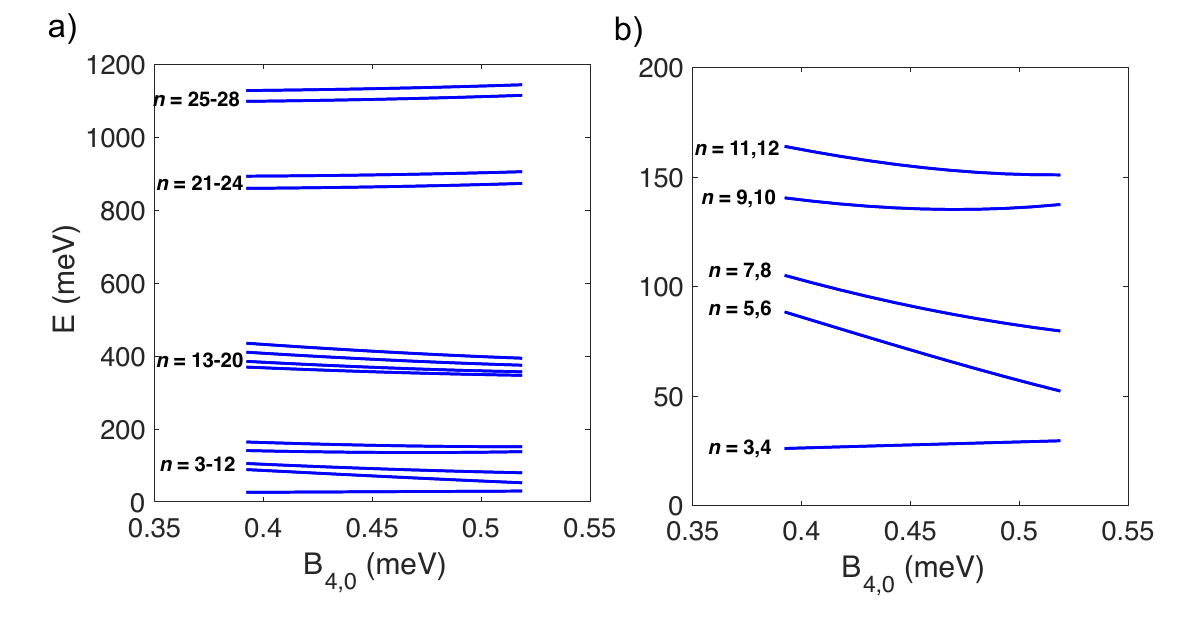}
\caption{ a) Energy eigenvalues as a function of $B_{4,0}$. The eigenstates are labelled by $n$. b) Expanded view of the low energy part of panel a).}

\label{fig:Elevels}
\end{figure}

Fits to these measured single-ion levels were carried out using a least squares minimization routine.  The dynamic structure factor at constant $|Q|$, i.e, $S(E)$, was calculated from the single-ion model with local trigonal disorder described in Section \ref{sec:single-ion}, and was compared to the $E_i = 250$ meV data averaged from $|Q| = 4-5$ \AA$^{-1}$ (Fig. \ref{fig:CEF} c)).  The measured energies and uncertainties of the higher excited levels ($364 \pm 1$ meV and $908 \pm 30$ meV) were included as constraints on the minimization, but no attempt was made to compare their relative intensities, which are difficult to accurately determine due to instrumental effects.  The HWHM parameters $\Gamma_{\langle n,n' \rangle}$ in Eqn. \ref{eqn:SofE} were held fixed to the calculated instrumental resolution for most values of $n$ and $n'$, but were increased to 13.5 meV for $n = 1,2$ and $n'=3,4$ to account for the dispersion of the first excited level.   The coherent and incoherent nuclear elastic scattering was modeled by a Gaussian with FWHM fixed to the instrumental resolution ($\delta E = 12.7$ meV), and the area of that elastic signal was used as an additional fitting parameter ($C_E$).   In total there were seven fitting parameters; the three average crystal field parameters ($B_{2,0}$, $B_{4,0}$, and $B_{4,3}$), a width of the linear distribution of $B_{4,0}$ ($\Delta B_{4,0}$),  the orbital reduction parameter $p$, the elastic line area $C_E$, and an overall scale factor $C$.  These fitted parameters and the associated average $g$-tensors with their percent variation are listed in Table \ref{tab:fitparams}.  The energy eigenvalues for the average $B_{l,m}$, their variation over the full $\Delta B_{4,0}$ range, as well as transition matrix elements, are listed in Table \ref{tab:eigenvalues}.  The energy levels as a function of $B_{4,0}$ are visualized in Figure \ref{fig:Elevels}. Note that due to the overall similarity between the single-ion levels of NCCF and NSCF, as well as the limited accuracy of the simple model of local disorder (as discussed in Section \ref{sec:discussion}), only one set of parameters are presented.  Although this fit is not likely to be unique in the sense that the broad observed transitions and minimal modeling of disorder prevent a precise determination of the parameters, the obtained parameters adequately capture the main features of the single-ion levels in both materials and provide the key physical insights that can be gained from these data.

Finally, the high temperature dc magnetic susceptibility can be compared to the measured susceptibility of NCCF from Ref. \onlinecite{krizan2014nacaco}, using the average $B_{l,m}$ parameters shown in Table \ref{tab:fitparams}.   Fitting the inverse susceptibility from $T=$100-300 K using Eqns. \ref{eqn:suscept} and \ref{eqn:chiMF}, the two parameters $\chi_{\text{dia}} = -3.1(1)\times10^{-4}$ emu/mol and $\lambda_W = 23.99(5)$ mol/emu were determined, and excellent agreement is found with the high temperature data (Fig. \ref{fig:CEF} f).  This agreement further confirms the validity of the single-ion scheme determined from the neutron scattering results.

\begin{table}[]
\centering
\begin{tabular}{|l|l|l}
\cline{1-2}
Parameter & Fitted value &                                                           \\ \cline{1-2}
$B_{2,0}$      	     &  25.8(5) (meV)          &                                                           \\ \cline{1-2}
$B_{4,0}$      	     &    0.455(5) (meV)      &                                                           \\ \cline{1-2}
$B_{4,3}$       	    &         14.01(7)  (meV)  &                                                           \\ \cline{1-2}
$\Delta B_{4,0}$	  &        0.127(4) (meV)  & \multirow{-5}{*}{}                                        \\ \hline
$p$         		  &         1.021(4)      	& \multicolumn{1}{l|}{\% variation} \\ \hline
$g_z$         	   &        1.87         & \multicolumn{1}{l|}{8.4 \%}                                     \\ \hline
$g_{xy}$         	   &        6.08         & \multicolumn{1}{l|}{ 4.1 \%}                                    \\ \hline
\end{tabular}
\caption{Parameters for the single-ion Hamiltonian, fit to inelastic neutron scattering data in NCCF.  The values of $g_z$ and $g_{xy}$ are those obtained from the mean value of the CEF parameters (i.e., when $\Delta B_{40} = 0$).}
\label{tab:fitparams}
\end{table}

\begin{table}

    \begin{tabular}{|l|l|l|l|}
    \hline

    State ($\nu_{n'}$) &  $E_{n'}$ for $\Delta B_{4,0} = 0$ & Range of $E_{n'}$ & $M^2_{\langle 1,n' \rangle}$+$M^2_{\langle 2,n' \rangle}$ \\ \hline
   $\nu_1$,$\nu_2$      	      	& 0.0              & 0.0            &   19.48         \\
 $\nu_3$,$\nu_4$           	& 27.7            & 3.5            & 17.06           \\
 $\nu_5$,$\nu_6$            	& 69.4              & 36.1            & 3.55          \\
 $\nu_7$,$\nu_8$           	& 90.0             & 25.4           & 0.37           \\
 $\nu_9$,$\nu_{10}$        	& 135.4              & 3.03            & 0.12           \\
 $\nu_{11}$,$\nu_{12}$           	& 154.3              & 13.0            & 0.04           \\
 $\nu_{13}$,$\nu_{14}$          & 355.6              & 22.6            & 0.51          \\
 $\nu_{15}$,$\nu_{16}$          & 367.6              & 28.6            & 0.67           \\
  $\nu_{17}$,$\nu_{18}$           & 389.3              & 35.6            & 0.83           \\
   $\nu_{19}$,$\nu_{20}$          & 411.1              & 41.1            & 0.17          \\
  $\nu_{21}$,$\nu_{22}$          & 863.9             & 13.5            & 0.11          \\
  $\nu_{23}$,$\nu_{24}$          & 896.7              & 12.2            & 0.71          \\
   $\nu_{25}$,$\nu_{26}$          & 1103.9        & 16.5            & 0.02           \\                        
   $\nu_{27}$,$\nu_{28}$          & 1133.3     & 15.8           & 0.01           \\ \hline
    \end{tabular}   
\caption{Calculated energy levels (in meV) from the fit shown in Fig. \ref{fig:CEF}, and the corresponding energy ranges arising from the variation in a crystal field parameter ($\Delta B_{4,0}$) which was used to model the local disorder at the Co$^{2+}$ site. The sum of the square of the transition matrix elements from the ground doublet to each excited doublet are listed, for the average CEF parameters ($M^2_{\langle n,n' \rangle} = \sum_\alpha |\langle \nu_n | \hat{L}_\alpha + 2 \hat{S}_\alpha | \nu_{n'} \rangle|^2$). Note that not all of the transitions listed in this table have been observed; refer to Fig. \ref{fig:CEF} for the correspondence between the measured and calculated peak positions.}
\label{tab:eigenvalues} 
\end{table}

\section{Discussion}
 \label{sec:discussion}

\subsubsection{Single-ion model} 
The single-ion fits suggest that the reduction of spin orbit coupling (quantified by $p$) is not significant, since it refines to 1.021(4), which is nearly 1.0.  This is in contrast to KCoF$_3$, for which $p = $0.93 \cite{buyers1971excitations}), indicating orbital moment reduction due to covalency.  Thus, naively, NCCF and NSCF seem to have strongly ionic bonds.

The model used to fit the single-ion levels, presented in Section \ref{sec:single-ion}, accounts for the local disorder in a minimal way.   The local symmetry at the Co$^{2+}$ site is assumed to remain trigonal, based on vibrational spectroscopy of the non-magnetic structural analog NaCaMg$_2$F$_7$ from Ref. \onlinecite{oliveira2004crystal}.  In that study it was found that the number and selection rules of vibrational modes was inconsistent with the full $Fd\bar{3}m$ symmetry of the average structure, but could be accounted for using the $F4\bar{3}m$ subgroup.  Using the online software package {\sc isodistort}\cite{campbell2006isodisplace}, and assuming atomic displacements consistent with this symmetry lowering, we find that the point group of the Co site remains trigonal, but is reduced from $D_{3d}$ to $C_{3v}$ (note that on average, these local distortions would cancel out and produce the $Fd\bar{3}m$ space group that is observed for the average structure).   This implies that only three CEF parameters are required, but that there may be some variation in their values from site to site, as the Co-F bond angles and distances (and therefore the CEF) are modified by the local disorder.   In the absence of a more realistic model for the distribution of local disorder, we initially allowed a linear variation in each of the crystal field parameters in Eqn. \ref{eqn:Hcef} and averaged the calculated $S(E)$ for each set of parameters.  We found that only a variation in $B_{4,0}$ (which we call $\Delta B_{4,0}$) was necessary to reproduce the observed widths and positions of the peaks observed via INS (Fig. \ref{fig:CEF}).  Specifically, the energies of the $\nu_5$, $\nu_6$ and  $\nu_7$, $\nu_8$ doublets are more widely distributed compared to the other levels, consistent with the broad band of magnetic intensity observed between $\sim$ 40 and 100 meV in INS, without creating excessive widening (beyond the measured widths) of the higher energy levels.  It may be noted that this model does not perfectly account for the intensities and positions of all features, particularly near the modes near 121 meV.  This is not unexpected, as the model we are using for the disorder is simplified.  The numerical values we report for the $g$-tensor and its variation ($g_z = 1.87 \pm 8.4$ \% and $g_{xy} = 6.08 \pm 4.1$\%) can nevertheless be used as good estimates.

 A more accurate model of the disorder could perhaps be developed based on additional experimental information about the local structure of these materials.  To this end, a $^{23}$Na NMR study of NCCF has revealed an unusual high temperature response suggesting that there are two main crystallographic environments for Na \cite{sarkar2016unconventional}.  So far no model to explain this observation has been proposed.  A measurement of the neutron or x-ray Pair Distribution Function (PDF) could provide additional information.
 
Although there remains some uncertainty in the distribution of $g$-tensor values, the results of our single-ion study unequivocally indicate that NSCF and NCCF are XY pyrochlores with well-isolated Kramers doublets.  This doublet nature of the low energy degrees of freedom is consistent with estimates of the spin entropy from specific heat which reach nearly $R$ln2 per mole of Co \cite{krizan2014nacaco,krizan2015nasrco2f7}, as well as with the sum rule analysis of low energy neutron scattering intensities \cite{ross2016static}.  Our results show that interactions in these materials may be treated within a pseudo-spin 1/2 model with XY effective exchange.  If the dominant interactions occur between nearest-neighbors, this would produce a model similar to the one used to describe Er$_2$Ti$_2$O$_7$ and Er$_2$Sn$_2$O$_7$ \cite{zhitomirsky2012quantum,savary2012order,guitteny2013palmer}.  The variation in the $g$-tensor implies that exchange disorder is needed in the effective model.

\subsubsection{Susceptibility}
The dc susceptibility in both NSCF and NCCF can be well-accounted for by the single-ion model using with the average $B_{l,m}$ parameters reported in Table \ref{tab:fitparams}.  The model naturally explains the high effective moments of these two materials ($\sim$ 6 $\mu_B$) as determined from Curie-Weiss fits of the inverse susceptibility over the range 100 K to 300 K \cite{krizan2014nacaco}.   Another feature of the model is a downturn in $1/\chi$ near 50 K; this experimentally observed feature was previously assumed to be due to ferromagnetic correlations, but it can now be understood as changes in the thermal population of excited single-ion levels.  The fitted Weiss molecular field constant, $\lambda_W$ = 23.99(4) mol/emu, confirms that AFM interactions are dominant, in agreement with the low energy INS results.    

\subsubsection{Comparison of \nccf \ and \nscf} 
The neutron scattering results presented in Section \ref{sec:results} indicate that the magnetic correlations and single ion properties of NSCF and NCCF are nearly identical.    NCCF and NSCF have freezing temperatures of 2.4 K and 3.0 K, respectively.  The increased freezing temperature of NSCF seems likely to be related to the larger differential ionic radius on the $A$ site: for 8-fold coordinated Na$^{+}$, $r =  1.18$ \AA, for Ca$^{2+}$ $r =  1.12$ \AA \ and for Sr$^{2+}$ $r =  1.26$ \cite{shannon1976revised}.  The bigger this size difference, the more severe the local disorder is expected to be.  The measured single-ion levels are very similar, with the only notable difference being the position of the first excited level, which is slightly higher in energy for NSCF compared to NCCF.  However, within the accuracy of our model, there is no discernible difference in the single-ion ground state wavefunctions.  Thus, it seems likely that both NSCF and NCCF could be treated within the same low energy effective model, but the strength of exchange disorder should be enhanced for NSCF.

 \subsubsection{Co$^{2+}$ pyrochlores} 
 
 Given the promise of extending the phenomenology of anisotropic rare earth pyrochlores to materials with higher interaction strengths, searching for other Co$^{2+}$-based pyrochlores seems desirable.  To the best of our knowledge, only one other material has been reported with a Co$^{2+}$ pyrochlore sublattice: the spinel GeCo$_2$O$_4$.    This material has been discussed in terms of its $S_{\text{eff}}=1/2$ single-ion ground state, and some unusual details of the higher energy excitation spectrum have been modeled in terms of molecular magnetism involving neighboring tetrahedra of these $S_{\text{eff}}=1/2$ moments \cite{tomiyasu2011molecular}.   In GeCo$_2$O$_4$ the CoO$_6$ octahedra are not distorted, and the moment is therefore isotropic. 
 
  An interesting question for future exploration is whether there exists a Co$^{2+}$-based \emph{spin ice} material (either quantum or classical).   Such a material would have the advantage that the emergent monopole dynamics would occur at temperatures much easier to access, possibly even high enough to be useful for applications.   If the U(1) quantum spin liquid phase of \emph{quantum} spin ice could be produced in a Co$^{2+}$ pyrochlore, its emergent photon modes would have a more accessible bandwidth compared to the rare earth-based candidates, making them more easily identifiable with low energy inelastic neutron scattering and thermodynamic measurements.  To achieve the necessary Ising anisotropy within a  Co$^{2+}$ pyrochlore, the octahedral environment would need to be \emph{elongated} along the $\langle 111 \rangle$ directions rather than compressed as it is for NCCF and NSCF.

\section{Conclusions}
 \label{sec:conclusions}
 
We have presented inelastic neutron scattering results on the pyrochlore materials \nccf \ and \nscf.  The low energy response of \nscf \ (below 10 meV) confirms that it hosts the same type of spin correlations as \nccf, which are fully dynamic above the freezing temperatures (2.4 - 3.0 K) but static and short range correlated below.  These correlations are well-described in terms of static and dynamic antiferromagnetic XY spin clusters, as discussed in Ref. \onlinecite{ross2016static}.   Below $T_F$, the XY configurations are frozen, short range correlated versions of the $\psi_2$ and $\psi_3$ states which are known to be selected through various types of order-by-disorder in the XY AFM models of the pyrochlore lattice.

Our main result is the measurement and analysis of  the single-ion levels of Co$^{2+}$ in these materials. These are well-modeled by an intermediate spin orbit coupling Hamiltonian.  We incorporated a distribution of crystal field parameters to account for local disorder brought about by the split non-magnetic $A$ site.   The ground state of Co$^{2+}$ in these environments is always a Kramers doublet with an XY $g$-tensor.  The average crystal field parameters produce $g_z = 1.87$ and $g_{xy} = 6.08$, with a variation of 8.4\% and 4.1\%, respectively.  The single-ion ground state doublets are separated from the first excited states by $\sim$300 K.   Thus, at low temperatures ($T<\theta_{CW} \sim150$ K), where spin correlations develop, a low energy effective theory built from $S=1/2$ operators can be used to describe these materials.  Due to the XY g-tensor and its variation, the effective exchange interactions are likely to be XY-like with bond (exchange) disorder.   A basic model of the AFM XY pyrochlore with exchange disorder has been predicted to lead to a long range ordered state.  However, \nccf \ and \nscf \ do not seem to conform to this prediction.

In summary, our results show that these materials can be thought of as ``high temperature'' versions of the $S_{\text{eff}} = 1/2$ AFM XY pyrochlore, which has been studied in relation to the rare earth oxide series of pyrochlores.  Their strong interactions allow the exploration of a wider temperature range of this model.  The inherent local disorder arising from the split non-magnetic $A$-site is likely to produce exchange disorder, in part through the local variations of the $g$-tensor that we have estimated here.
The role of this exchange disorder in a model appropriate to these materials requires more investigation, as the ground states appear to be contrary to the prediction that \emph{quenched} order by disorder selects long range order.  Finally,  \nccf \ and \nscf \ are more frustrated ($f \sim 50$) versions of the AFM XY pyrochlore than the canonical example, Er$_2$Ti$_2$O$_7$ ($f = 20$).  This may indicate that they are closer to a phase boundary in the general anisotropic exchange model developed for the pyrochlore lattice; this proximity would make the materials more sensitive to quantum fluctuations, and may even put them in proximity to a quantum disordered phase.

\begin{acknowledgments}
The authors gratefully acknowledges discussions with K. Kimura, the use of software adapted from work by G. Hester and J. O'Brien, and the encouragement of C. L. Broholm. The neutron scattering data were reduced using the Mantid \cite{arnold2014mantid} and DAVE\cite{azuah2009dave} software packages.  The crystal growth activities at Princeton were supported by the US Department of Energy, office of Basic Energy Sciences, Division of Material Sciences and Engineering under grant DE-FG02-08ER46544.  This work utilized neutron scattering facilities supported in part by the National Science Foundation under Agreement No. DMR-1508249.  Research using ORNL's Spallation Neutron Source was sponsored by the Scientific User Facilities Division, Office of Basic Energy Sciences, U.S. Department of Energy.  
\end{acknowledgments}

\appendix

\section{$Q$-dependence of modes observed with $E_i$ = 700 meV and 250 meV}
\label{sec:Qdep}

\begin{figure}[!h]  
\centering
\includegraphics[ width=\columnwidth]{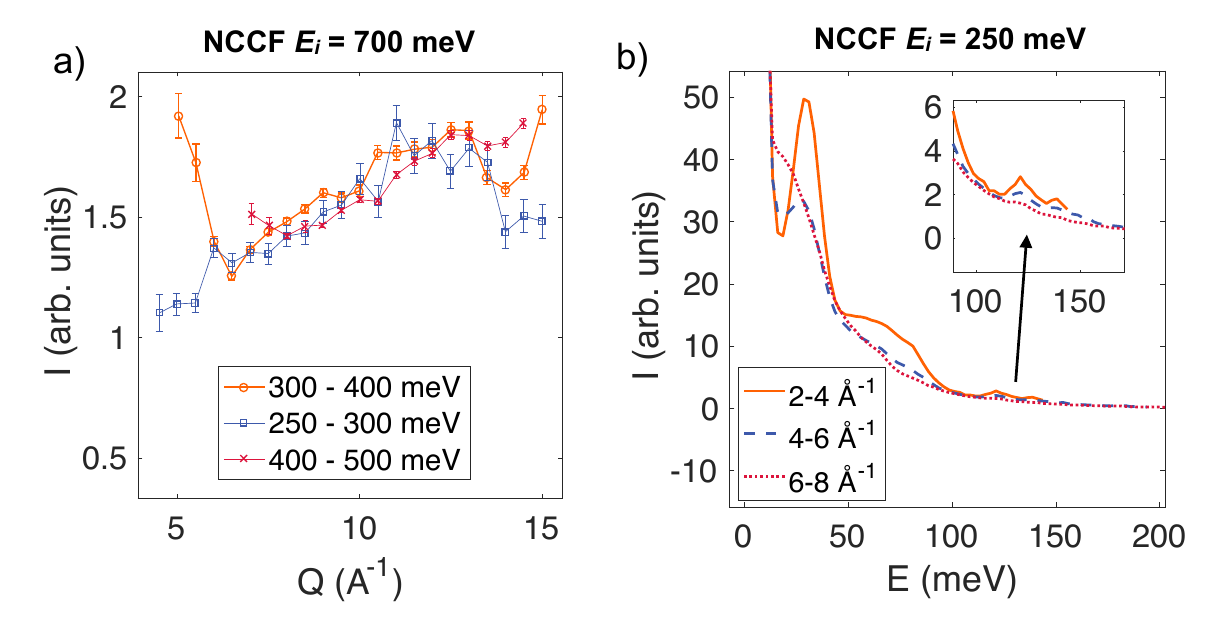}
\caption{a) $Q$ dependence of intensity for energy ranges spanning the identified single-ion mode at $\sim$ 360 meV (300 to 400 meV), just above the single-ion mode (400 to 500 meV, intensity shifted up for comparison), and just below the single-ion mode (250-300 meV, intensity shifted down for comparison).  b) Energy dependence of neutron scattering intensity in different $Q$ ranges for $E_i = $250 meV.}
\label{fig:Qdep}
\end{figure}

The inelastic modes shown in Figure \ref{fig:CEF}  a)-d) can be identified as being magnetic in origin due to the $|Q|$-dependence of their intensities.  The intensity of magnetic features (such as single-ion levels) should decrease with increasing $|Q|$ due to the magnetic form factor \cite{lovesey1984theory}.  Meanwhile the intensity of features arising from phonons (whether single- or multi-phonon processes) should show the opposite trend.  For the mode identified near 360 meV in NCCF (using $E_i$ = 700 meV), the $|Q$-dependence of intensity reveals a combination of both magnetic and multi-phonon contributions.  This is demonstrated in Figure \ref{fig:Qdep} a), which compares intensity near 360 meV to that in energy ranges just above and just below the observed mode.  This demonstrates that the peak observed near 360 meV is magnetic in origin.  The intensity of the features observed using $E_i$ = 250 meV are also shown for three different $Q$ ranges in panel b), demonstrating their magnetic nature.

%\bibliography{../../NaCaCo2F7_bib_jointpaper.bib}

\begin{thebibliography}{59}%
\makeatletter
\providecommand \@ifxundefined [1]{%
 \@ifx{#1\undefined}
}%
\providecommand \@ifnum [1]{%
 \ifnum #1\expandafter \@firstoftwo
 \else \expandafter \@secondoftwo
 \fi
}%
\providecommand \@ifx [1]{%
 \ifx #1\expandafter \@firstoftwo
 \else \expandafter \@secondoftwo
 \fi
}%
\providecommand \natexlab [1]{#1}%
\providecommand \enquote  [1]{``#1''}%
\providecommand \bibnamefont  [1]{#1}%
\providecommand \bibfnamefont [1]{#1}%
\providecommand \citenamefont [1]{#1}%
\providecommand \href@noop [0]{\@secondoftwo}%
\providecommand \href [0]{\begingroup \@sanitize@url \@href}%
\providecommand \@href[1]{\@@startlink{#1}\@@href}%
\providecommand \@@href[1]{\endgroup#1\@@endlink}%
\providecommand \@sanitize@url [0]{\catcode `\\12\catcode `\$12\catcode
  `\&12\catcode `\#12\catcode `\^12\catcode `\_12\catcode `\%12\relax}%
\providecommand \@@startlink[1]{}%
\providecommand \@@endlink[0]{}%
\providecommand \url  [0]{\begingroup\@sanitize@url \@url }%
\providecommand \@url [1]{\endgroup\@href {#1}{\urlprefix }}%
\providecommand \urlprefix  [0]{URL }%
\providecommand \Eprint [0]{\href }%
\providecommand \doibase [0]{http://dx.doi.org/}%
\providecommand \selectlanguage [0]{\@gobble}%
\providecommand \bibinfo  [0]{\@secondoftwo}%
\providecommand \bibfield  [0]{\@secondoftwo}%
\providecommand \translation [1]{[#1]}%
\providecommand \BibitemOpen [0]{}%
\providecommand \bibitemStop [0]{}%
\providecommand \bibitemNoStop [0]{.\EOS\space}%
\providecommand \EOS [0]{\spacefactor3000\relax}%
\providecommand \BibitemShut  [1]{\csname bibitem#1\endcsname}%
\let\auto@bib@innerbib\@empty
%</preamble>
\bibitem [{\citenamefont {Ross}\ \emph {et~al.}(2016)\citenamefont {Ross},
  \citenamefont {Krizan}, \citenamefont {Rodriguez-Rivera}, \citenamefont
  {Cava},\ and\ \citenamefont {Broholm}}]{ross2016static}%
  \BibitemOpen
  \bibfield  {author} {\bibinfo {author} {\bibfnamefont {K.~A.}\ \bibnamefont
  {Ross}}, \bibinfo {author} {\bibfnamefont {J.~W.}\ \bibnamefont {Krizan}},
  \bibinfo {author} {\bibfnamefont {J.~A.}\ \bibnamefont {Rodriguez-Rivera}},
  \bibinfo {author} {\bibfnamefont {R.~J.}\ \bibnamefont {Cava}}, \ and\
  \bibinfo {author} {\bibfnamefont {C.~L.}\ \bibnamefont {Broholm}},\
  }\href@noop {} {\bibfield  {journal} {\bibinfo  {journal} {Physical Review
  B}\ }\textbf {\bibinfo {volume} {93}},\ \bibinfo {pages} {014433} (\bibinfo
  {year} {2016})}\BibitemShut {NoStop}%
\bibitem [{\citenamefont {Gardner}\ \emph {et~al.}(2010)\citenamefont
  {Gardner}, \citenamefont {Gingras},\ and\ \citenamefont
  {Greedan}}]{gardner2010magnetic}%
  \BibitemOpen
  \bibfield  {author} {\bibinfo {author} {\bibfnamefont {J.~S.}\ \bibnamefont
  {Gardner}}, \bibinfo {author} {\bibfnamefont {M.~J.~P.}\ \bibnamefont
  {Gingras}}, \ and\ \bibinfo {author} {\bibfnamefont {J.~E.}\ \bibnamefont
  {Greedan}},\ }\href@noop {} {\bibfield  {journal} {\bibinfo  {journal}
  {Reviews of Modern Physics}\ }\textbf {\bibinfo {volume} {82}},\ \bibinfo
  {pages} {53} (\bibinfo {year} {2010})}\BibitemShut {NoStop}%
\bibitem [{\citenamefont {Shannon}\ and\ \citenamefont
  {Sleight}(1968)}]{shannon1968synthesis}%
  \BibitemOpen
  \bibfield  {author} {\bibinfo {author} {\bibfnamefont {R.~D.}\ \bibnamefont
  {Shannon}}\ and\ \bibinfo {author} {\bibfnamefont {A.~W.}\ \bibnamefont
  {Sleight}},\ }\href@noop {} {\bibfield  {journal} {\bibinfo  {journal}
  {Inorganic Chemistry}\ }\textbf {\bibinfo {volume} {7}},\ \bibinfo {pages}
  {1649} (\bibinfo {year} {1968})}\BibitemShut {NoStop}%
\bibitem [{\citenamefont {Dun}\ \emph {et~al.}(2015)\citenamefont {Dun},
  \citenamefont {Li}, \citenamefont {Freitas}, \citenamefont {Arrighi},
  \citenamefont {Cruz}, \citenamefont {Lee}, \citenamefont {Choi},
  \citenamefont {Cao}, \citenamefont {Silverstein}, \citenamefont {Wiebe} \emph
  {et~al.}}]{dun2015antiferromagnetic}%
  \BibitemOpen
  \bibfield  {author} {\bibinfo {author} {\bibfnamefont {Z.~L.}\ \bibnamefont
  {Dun}}, \bibinfo {author} {\bibfnamefont {X.}~\bibnamefont {Li}}, \bibinfo
  {author} {\bibfnamefont {R.~S.}\ \bibnamefont {Freitas}}, \bibinfo {author}
  {\bibfnamefont {E.}~\bibnamefont {Arrighi}}, \bibinfo {author} {\bibfnamefont
  {C.~R.~D.}\ \bibnamefont {Cruz}}, \bibinfo {author} {\bibfnamefont
  {M.}~\bibnamefont {Lee}}, \bibinfo {author} {\bibfnamefont {E.~S.}\
  \bibnamefont {Choi}}, \bibinfo {author} {\bibfnamefont {H.~B.}\ \bibnamefont
  {Cao}}, \bibinfo {author} {\bibfnamefont {H.~J.}\ \bibnamefont
  {Silverstein}}, \bibinfo {author} {\bibfnamefont {C.~R.}\ \bibnamefont
  {Wiebe}},  \emph {et~al.},\ }\href@noop {} {\bibfield  {journal} {\bibinfo
  {journal} {Physical Review B}\ }\textbf {\bibinfo {volume} {92}},\ \bibinfo
  {pages} {140407} (\bibinfo {year} {2015})}\BibitemShut {NoStop}%
\bibitem [{\citenamefont {Cai}\ \emph {et~al.}(2016)\citenamefont {Cai},
  \citenamefont {Cui}, \citenamefont {Li}, \citenamefont {Dun}, \citenamefont
  {Ma}, \citenamefont {dela Cruz}, \citenamefont {Jiao}, \citenamefont {Liao},
  \citenamefont {Sun}, \citenamefont {Li} \emph {et~al.}}]{cai2016high}%
  \BibitemOpen
  \bibfield  {author} {\bibinfo {author} {\bibfnamefont {Y.~Q.}\ \bibnamefont
  {Cai}}, \bibinfo {author} {\bibfnamefont {Q.}~\bibnamefont {Cui}}, \bibinfo
  {author} {\bibfnamefont {X.}~\bibnamefont {Li}}, \bibinfo {author}
  {\bibfnamefont {Z.}~\bibnamefont {Dun}}, \bibinfo {author} {\bibfnamefont
  {J.}~\bibnamefont {Ma}}, \bibinfo {author} {\bibfnamefont {C.}~\bibnamefont
  {dela Cruz}}, \bibinfo {author} {\bibfnamefont {Y.}~\bibnamefont {Jiao}},
  \bibinfo {author} {\bibfnamefont {J.}~\bibnamefont {Liao}}, \bibinfo {author}
  {\bibfnamefont {P.}~\bibnamefont {Sun}}, \bibinfo {author} {\bibfnamefont
  {Y.~Q.}\ \bibnamefont {Li}},  \emph {et~al.},\ }\href@noop {} {\bibfield
  {journal} {\bibinfo  {journal} {Physical Review B}\ }\textbf {\bibinfo
  {volume} {93}},\ \bibinfo {pages} {014443} (\bibinfo {year}
  {2016})}\BibitemShut {NoStop}%
\bibitem [{\citenamefont {Castelnovo}\ \emph {et~al.}(2012)\citenamefont
  {Castelnovo}, \citenamefont {Moessner},\ and\ \citenamefont
  {Sondhi}}]{castelnovo2012spin}%
  \BibitemOpen
  \bibfield  {author} {\bibinfo {author} {\bibfnamefont {C.}~\bibnamefont
  {Castelnovo}}, \bibinfo {author} {\bibfnamefont {R.}~\bibnamefont
  {Moessner}}, \ and\ \bibinfo {author} {\bibfnamefont {S.~L.}\ \bibnamefont
  {Sondhi}},\ }\href {\doibase 10.1146/annurev-conmatphys-020911-125058}
  {\bibfield  {journal} {\bibinfo  {journal} {Annual Review of Condensed Matter
  Physics}\ }\textbf {\bibinfo {volume} {3}},\ \bibinfo {pages} {35} (\bibinfo
  {year} {2012})}\BibitemShut {NoStop}%
\bibitem [{\citenamefont {Bramwell}\ \emph {et~al.}(1994)\citenamefont
  {Bramwell}, \citenamefont {Gingras},\ and\ \citenamefont
  {Reimers}}]{bramwell1994order}%
  \BibitemOpen
  \bibfield  {author} {\bibinfo {author} {\bibfnamefont {S.~T.}\ \bibnamefont
  {Bramwell}}, \bibinfo {author} {\bibfnamefont {M.~J.~P.}\ \bibnamefont
  {Gingras}}, \ and\ \bibinfo {author} {\bibfnamefont {J.~N.}\ \bibnamefont
  {Reimers}},\ }\href@noop {} {\bibfield  {journal} {\bibinfo  {journal}
  {Journal of Applied Physics}\ }\textbf {\bibinfo {volume} {75}},\ \bibinfo
  {pages} {5523} (\bibinfo {year} {1994})}\BibitemShut {NoStop}%
\bibitem [{\citenamefont {Wong}\ \emph {et~al.}(2013)\citenamefont {Wong},
  \citenamefont {Hao},\ and\ \citenamefont {Gingras}}]{wong2013ground}%
  \BibitemOpen
  \bibfield  {author} {\bibinfo {author} {\bibfnamefont {A.~W.~C.}\
  \bibnamefont {Wong}}, \bibinfo {author} {\bibfnamefont {Z.}~\bibnamefont
  {Hao}}, \ and\ \bibinfo {author} {\bibfnamefont {M.~J.~P.}\ \bibnamefont
  {Gingras}},\ }\href@noop {} {\bibfield  {journal} {\bibinfo  {journal}
  {Physical Review B}\ }\textbf {\bibinfo {volume} {88}},\ \bibinfo {pages}
  {144402} (\bibinfo {year} {2013})}\BibitemShut {NoStop}%
\bibitem [{\citenamefont {McClarty}\ \emph {et~al.}(2014)\citenamefont
  {McClarty}, \citenamefont {Stasiak},\ and\ \citenamefont
  {Gingras}}]{mcclarty2014order}%
  \BibitemOpen
  \bibfield  {author} {\bibinfo {author} {\bibfnamefont {P.~A.}\ \bibnamefont
  {McClarty}}, \bibinfo {author} {\bibfnamefont {P.}~\bibnamefont {Stasiak}}, \
  and\ \bibinfo {author} {\bibfnamefont {M.~J.~P.}\ \bibnamefont {Gingras}},\
  }\href@noop {} {\bibfield  {journal} {\bibinfo  {journal} {Physical Review
  B}\ }\textbf {\bibinfo {volume} {89}},\ \bibinfo {pages} {024425} (\bibinfo
  {year} {2014})}\BibitemShut {NoStop}%
\bibitem [{\citenamefont {Champion}\ \emph {et~al.}(2003)\citenamefont
  {Champion}, \citenamefont {Harris}, \citenamefont {Holdsworth}, \citenamefont
  {Wills}, \citenamefont {Balakrishnan}, \citenamefont {Bramwell},
  \citenamefont {{\v{C}}i{\v{z}}m{\'a}r}, \citenamefont {Fennell},
  \citenamefont {Gardner}, \citenamefont {Lago} \emph
  {et~al.}}]{champion2003er}%
  \BibitemOpen
  \bibfield  {author} {\bibinfo {author} {\bibfnamefont {J.~D.~M.}\
  \bibnamefont {Champion}}, \bibinfo {author} {\bibfnamefont {M.~J.}\
  \bibnamefont {Harris}}, \bibinfo {author} {\bibfnamefont {P.~C.~W.}\
  \bibnamefont {Holdsworth}}, \bibinfo {author} {\bibfnamefont {A.~S.}\
  \bibnamefont {Wills}}, \bibinfo {author} {\bibfnamefont {G.}~\bibnamefont
  {Balakrishnan}}, \bibinfo {author} {\bibfnamefont {S.~T.}\ \bibnamefont
  {Bramwell}}, \bibinfo {author} {\bibfnamefont {E.}~\bibnamefont
  {{\v{C}}i{\v{z}}m{\'a}r}}, \bibinfo {author} {\bibfnamefont {T.}~\bibnamefont
  {Fennell}}, \bibinfo {author} {\bibfnamefont {J.~S.}\ \bibnamefont
  {Gardner}}, \bibinfo {author} {\bibfnamefont {J.}~\bibnamefont {Lago}},
  \emph {et~al.},\ }\href@noop {} {\bibfield  {journal} {\bibinfo  {journal}
  {Physical Review B}\ }\textbf {\bibinfo {volume} {68}},\ \bibinfo {pages}
  {020401} (\bibinfo {year} {2003})}\BibitemShut {NoStop}%
\bibitem [{\citenamefont {Zhitomirsky}\ \emph {et~al.}(2012)\citenamefont
  {Zhitomirsky}, \citenamefont {Gvozdikova}, \citenamefont {Holdsworth},\ and\
  \citenamefont {Moessner}}]{zhitomirsky2012quantum}%
  \BibitemOpen
  \bibfield  {author} {\bibinfo {author} {\bibfnamefont {M.~E.}\ \bibnamefont
  {Zhitomirsky}}, \bibinfo {author} {\bibfnamefont {M.~V.}\ \bibnamefont
  {Gvozdikova}}, \bibinfo {author} {\bibfnamefont {P.~C.~W.}\ \bibnamefont
  {Holdsworth}}, \ and\ \bibinfo {author} {\bibfnamefont {R.}~\bibnamefont
  {Moessner}},\ }\href@noop {} {\bibfield  {journal} {\bibinfo  {journal}
  {Physical Review Letters}\ }\textbf {\bibinfo {volume} {109}},\ \bibinfo
  {pages} {077204} (\bibinfo {year} {2012})}\BibitemShut {NoStop}%
\bibitem [{\citenamefont {Savary}\ \emph {et~al.}(2012)\citenamefont {Savary},
  \citenamefont {Ross}, \citenamefont {Gaulin}, \citenamefont {Ruff},\ and\
  \citenamefont {Balents}}]{savary2012order}%
  \BibitemOpen
  \bibfield  {author} {\bibinfo {author} {\bibfnamefont {L.}~\bibnamefont
  {Savary}}, \bibinfo {author} {\bibfnamefont {K.~A.}\ \bibnamefont {Ross}},
  \bibinfo {author} {\bibfnamefont {B.~D.}\ \bibnamefont {Gaulin}}, \bibinfo
  {author} {\bibfnamefont {J.~P.}\ \bibnamefont {Ruff}}, \ and\ \bibinfo
  {author} {\bibfnamefont {L.}~\bibnamefont {Balents}},\ }\href@noop {}
  {\bibfield  {journal} {\bibinfo  {journal} {Physical Review Letters}\
  }\textbf {\bibinfo {volume} {109}},\ \bibinfo {pages} {167201} (\bibinfo
  {year} {2012})}\BibitemShut {NoStop}%
\bibitem [{\citenamefont {Oitmaa}\ \emph {et~al.}(2013)\citenamefont {Oitmaa},
  \citenamefont {Singh}, \citenamefont {Javanparast}, \citenamefont {Day},
  \citenamefont {Bagheri},\ and\ \citenamefont {Gingras}}]{oitmaa2013phase}%
  \BibitemOpen
  \bibfield  {author} {\bibinfo {author} {\bibfnamefont {J.}~\bibnamefont
  {Oitmaa}}, \bibinfo {author} {\bibfnamefont {R.~R.~P.}\ \bibnamefont
  {Singh}}, \bibinfo {author} {\bibfnamefont {B.}~\bibnamefont {Javanparast}},
  \bibinfo {author} {\bibfnamefont {A.~G.~R.}\ \bibnamefont {Day}}, \bibinfo
  {author} {\bibfnamefont {B.~V.}\ \bibnamefont {Bagheri}}, \ and\ \bibinfo
  {author} {\bibfnamefont {M.~J.~P.}\ \bibnamefont {Gingras}},\ }\href@noop {}
  {\bibfield  {journal} {\bibinfo  {journal} {Physical Review B}\ }\textbf
  {\bibinfo {volume} {88}},\ \bibinfo {pages} {220404} (\bibinfo {year}
  {2013})}\BibitemShut {NoStop}%
\bibitem [{\citenamefont {Maryasin}\ and\ \citenamefont
  {Zhitomirsky}(2014)}]{maryasin2014order}%
  \BibitemOpen
  \bibfield  {author} {\bibinfo {author} {\bibfnamefont {V.~S.}\ \bibnamefont
  {Maryasin}}\ and\ \bibinfo {author} {\bibfnamefont {M.~E.}\ \bibnamefont
  {Zhitomirsky}},\ }\href@noop {} {\bibfield  {journal} {\bibinfo  {journal}
  {Physical Review B}\ }\textbf {\bibinfo {volume} {90}},\ \bibinfo {pages}
  {094412} (\bibinfo {year} {2014})}\BibitemShut {NoStop}%
\bibitem [{\citenamefont {Andreanov}\ and\ \citenamefont
  {McClarty}(2015)}]{andreanov2015order}%
  \BibitemOpen
  \bibfield  {author} {\bibinfo {author} {\bibfnamefont {A.}~\bibnamefont
  {Andreanov}}\ and\ \bibinfo {author} {\bibfnamefont {P.~A.}\ \bibnamefont
  {McClarty}},\ }\href@noop {} {\bibfield  {journal} {\bibinfo  {journal}
  {Physical Review B}\ }\textbf {\bibinfo {volume} {91}},\ \bibinfo {pages}
  {064401} (\bibinfo {year} {2015})}\BibitemShut {NoStop}%
\bibitem [{\citenamefont {Ross}\ \emph {et~al.}(2012)\citenamefont {Ross},
  \citenamefont {Proffen}, \citenamefont {Dabkowska}, \citenamefont {Quilliam},
  \citenamefont {Yaraskavitch}, \citenamefont {Kycia},\ and\ \citenamefont
  {Gaulin}}]{ross2012lightly}%
  \BibitemOpen
  \bibfield  {author} {\bibinfo {author} {\bibfnamefont {K.~A.}\ \bibnamefont
  {Ross}}, \bibinfo {author} {\bibfnamefont {T.}~\bibnamefont {Proffen}},
  \bibinfo {author} {\bibfnamefont {H.~A.}\ \bibnamefont {Dabkowska}}, \bibinfo
  {author} {\bibfnamefont {J.~A.}\ \bibnamefont {Quilliam}}, \bibinfo {author}
  {\bibfnamefont {L.~R.}\ \bibnamefont {Yaraskavitch}}, \bibinfo {author}
  {\bibfnamefont {J.~B.}\ \bibnamefont {Kycia}}, \ and\ \bibinfo {author}
  {\bibfnamefont {B.~D.}\ \bibnamefont {Gaulin}},\ }\href@noop {} {\bibfield
  {journal} {\bibinfo  {journal} {Physical Review B}\ }\textbf {\bibinfo
  {volume} {86}},\ \bibinfo {pages} {174424} (\bibinfo {year}
  {2012})}\BibitemShut {NoStop}%
\bibitem [{\citenamefont {Yaouanc}\ \emph {et~al.}(2011)\citenamefont
  {Yaouanc}, \citenamefont {de~R{\'e}otier}, \citenamefont {Marin},\ and\
  \citenamefont {Glazkov}}]{yaouanc2011single}%
  \BibitemOpen
  \bibfield  {author} {\bibinfo {author} {\bibfnamefont {A.}~\bibnamefont
  {Yaouanc}}, \bibinfo {author} {\bibfnamefont {P.~D.}\ \bibnamefont
  {de~R{\'e}otier}}, \bibinfo {author} {\bibfnamefont {C.}~\bibnamefont
  {Marin}}, \ and\ \bibinfo {author} {\bibfnamefont {V.}~\bibnamefont
  {Glazkov}},\ }\href@noop {} {\bibfield  {journal} {\bibinfo  {journal}
  {Physical Review B}\ }\textbf {\bibinfo {volume} {84}},\ \bibinfo {pages}
  {172408} (\bibinfo {year} {2011})}\BibitemShut {NoStop}%
\bibitem [{\citenamefont {Taniguchi}\ \emph {et~al.}(2013)\citenamefont
  {Taniguchi}, \citenamefont {Kadowaki}, \citenamefont {Takatsu}, \citenamefont
  {F{\aa}k}, \citenamefont {Ollivier}, \citenamefont {Yamazaki}, \citenamefont
  {Sato}, \citenamefont {Yoshizawa}, \citenamefont {Shimura}, \citenamefont
  {Sakakibara} \emph {et~al.}}]{taniguchi2013long}%
  \BibitemOpen
  \bibfield  {author} {\bibinfo {author} {\bibfnamefont {T.}~\bibnamefont
  {Taniguchi}}, \bibinfo {author} {\bibfnamefont {H.}~\bibnamefont {Kadowaki}},
  \bibinfo {author} {\bibfnamefont {H.}~\bibnamefont {Takatsu}}, \bibinfo
  {author} {\bibfnamefont {B.}~\bibnamefont {F{\aa}k}}, \bibinfo {author}
  {\bibfnamefont {J.}~\bibnamefont {Ollivier}}, \bibinfo {author}
  {\bibfnamefont {T.}~\bibnamefont {Yamazaki}}, \bibinfo {author}
  {\bibfnamefont {T.~J.}\ \bibnamefont {Sato}}, \bibinfo {author}
  {\bibfnamefont {H.}~\bibnamefont {Yoshizawa}}, \bibinfo {author}
  {\bibfnamefont {Y.}~\bibnamefont {Shimura}}, \bibinfo {author} {\bibfnamefont
  {T.}~\bibnamefont {Sakakibara}},  \emph {et~al.},\ }\href@noop {} {\bibfield
  {journal} {\bibinfo  {journal} {Physical Review B}\ }\textbf {\bibinfo
  {volume} {87}},\ \bibinfo {pages} {060408} (\bibinfo {year}
  {2013})}\BibitemShut {NoStop}%
\bibitem [{\citenamefont {Krizan}\ and\ \citenamefont
  {Cava}(2014)}]{krizan2014nacaco}%
  \BibitemOpen
  \bibfield  {author} {\bibinfo {author} {\bibfnamefont {J.~W.}\ \bibnamefont
  {Krizan}}\ and\ \bibinfo {author} {\bibfnamefont {R.~J.}\ \bibnamefont
  {Cava}},\ }\href@noop {} {\bibfield  {journal} {\bibinfo  {journal} {Physical
  Review B}\ }\textbf {\bibinfo {volume} {89}},\ \bibinfo {pages} {214401}
  (\bibinfo {year} {2014})}\BibitemShut {NoStop}%
\bibitem [{\citenamefont {Krizan}\ and\ \citenamefont
  {Cava}(2015{\natexlab{a}})}]{krizan2015nasrco2f7}%
  \BibitemOpen
  \bibfield  {author} {\bibinfo {author} {\bibfnamefont {J.~W.}\ \bibnamefont
  {Krizan}}\ and\ \bibinfo {author} {\bibfnamefont {R.~J.}\ \bibnamefont
  {Cava}},\ }\href@noop {} {\bibfield  {journal} {\bibinfo  {journal} {Journal
  of Physics: Condensed Matter}\ }\textbf {\bibinfo {volume} {27}},\ \bibinfo
  {pages} {296002} (\bibinfo {year} {2015}{\natexlab{a}})}\BibitemShut
  {NoStop}%
\bibitem [{\citenamefont {Krizan}\ and\ \citenamefont
  {Cava}(2015{\natexlab{b}})}]{krizan2015nacan}%
  \BibitemOpen
  \bibfield  {author} {\bibinfo {author} {\bibfnamefont {J.~W.}\ \bibnamefont
  {Krizan}}\ and\ \bibinfo {author} {\bibfnamefont {R.~J.}\ \bibnamefont
  {Cava}},\ }\href@noop {} {\bibfield  {journal} {\bibinfo  {journal} {Physical
  Review B}\ }\textbf {\bibinfo {volume} {92}},\ \bibinfo {pages} {014406}
  (\bibinfo {year} {2015}{\natexlab{b}})}\BibitemShut {NoStop}%
\bibitem [{\citenamefont {Sanders}\ \emph {et~al.}(2016)\citenamefont
  {Sanders}, \citenamefont {Krizan}, \citenamefont {Plumb}, \citenamefont
  {McQueen},\ and\ \citenamefont {Cava}}]{sanders2016nasrmn2f7}%
  \BibitemOpen
  \bibfield  {author} {\bibinfo {author} {\bibfnamefont {M.~B.}\ \bibnamefont
  {Sanders}}, \bibinfo {author} {\bibfnamefont {J.~W.}\ \bibnamefont {Krizan}},
  \bibinfo {author} {\bibfnamefont {K.~W.}\ \bibnamefont {Plumb}}, \bibinfo
  {author} {\bibfnamefont {T.~M.}\ \bibnamefont {McQueen}}, \ and\ \bibinfo
  {author} {\bibfnamefont {R.~J.}\ \bibnamefont {Cava}},\ }\href@noop {}
  {\bibfield  {journal} {\bibinfo  {journal} {Journal of Physics: Condensed
  Matter}\ }\textbf {\bibinfo {volume} {29}},\ \bibinfo {pages} {045801}
  (\bibinfo {year} {2016})}\BibitemShut {NoStop}%
\bibitem [{\citenamefont {Zinkin}\ \emph {et~al.}(1997)\citenamefont {Zinkin},
  \citenamefont {Harris},\ and\ \citenamefont {Zeiske}}]{zinkin1997short}%
  \BibitemOpen
  \bibfield  {author} {\bibinfo {author} {\bibfnamefont {M.~P.}\ \bibnamefont
  {Zinkin}}, \bibinfo {author} {\bibfnamefont {M.~J.}\ \bibnamefont {Harris}},
  \ and\ \bibinfo {author} {\bibfnamefont {T.}~\bibnamefont {Zeiske}},\
  }\href@noop {} {\bibfield  {journal} {\bibinfo  {journal} {Physical Review
  B}\ }\textbf {\bibinfo {volume} {56}},\ \bibinfo {pages} {11786} (\bibinfo
  {year} {1997})}\BibitemShut {NoStop}%
\bibitem [{\citenamefont {Bl{\"o}te}\ \emph {et~al.}(1969)\citenamefont
  {Bl{\"o}te}, \citenamefont {Wielinga},\ and\ \citenamefont
  {Huiskamp}}]{blote1969heat}%
  \BibitemOpen
  \bibfield  {author} {\bibinfo {author} {\bibfnamefont {H.}~\bibnamefont
  {Bl{\"o}te}}, \bibinfo {author} {\bibfnamefont {R.}~\bibnamefont {Wielinga}},
  \ and\ \bibinfo {author} {\bibfnamefont {W.}~\bibnamefont {Huiskamp}},\
  }\href@noop {} {\bibfield  {journal} {\bibinfo  {journal} {Physica}\ }\textbf
  {\bibinfo {volume} {43}},\ \bibinfo {pages} {549} (\bibinfo {year}
  {1969})}\BibitemShut {NoStop}%
\bibitem [{\citenamefont {Hodges}\ \emph {et~al.}(2011)\citenamefont {Hodges},
  \citenamefont {de~R{\'e}otier}, \citenamefont {Yaouanc}, \citenamefont
  {Gubbens}, \citenamefont {King},\ and\ \citenamefont
  {Baines}}]{hodges2011magnetic}%
  \BibitemOpen
  \bibfield  {author} {\bibinfo {author} {\bibfnamefont {J.~A.}\ \bibnamefont
  {Hodges}}, \bibinfo {author} {\bibfnamefont {P.~D.}\ \bibnamefont
  {de~R{\'e}otier}}, \bibinfo {author} {\bibfnamefont {A.}~\bibnamefont
  {Yaouanc}}, \bibinfo {author} {\bibfnamefont {P.~C.~M.}\ \bibnamefont
  {Gubbens}}, \bibinfo {author} {\bibfnamefont {P.~J.~C.}\ \bibnamefont
  {King}}, \ and\ \bibinfo {author} {\bibfnamefont {C.}~\bibnamefont
  {Baines}},\ }\href@noop {} {\bibfield  {journal} {\bibinfo  {journal}
  {Journal of Physics: Condensed Matter}\ }\textbf {\bibinfo {volume} {23}},\
  \bibinfo {pages} {164217} (\bibinfo {year} {2011})}\BibitemShut {NoStop}%
\bibitem [{\citenamefont {Buyers}\ \emph {et~al.}(1971)\citenamefont {Buyers},
  \citenamefont {Holden}, \citenamefont {Svensson}, \citenamefont {Cowley},\
  and\ \citenamefont {Hutchings}}]{buyers1971excitations}%
  \BibitemOpen
  \bibfield  {author} {\bibinfo {author} {\bibfnamefont {W.~J.~L.}\
  \bibnamefont {Buyers}}, \bibinfo {author} {\bibfnamefont {T.~M.}\
  \bibnamefont {Holden}}, \bibinfo {author} {\bibfnamefont {E.~C.}\
  \bibnamefont {Svensson}}, \bibinfo {author} {\bibfnamefont {R.~A.}\
  \bibnamefont {Cowley}}, \ and\ \bibinfo {author} {\bibfnamefont {M.~T.}\
  \bibnamefont {Hutchings}},\ }\href@noop {} {\bibfield  {journal} {\bibinfo
  {journal} {Journal of Physics C: Solid State Physics}\ }\textbf {\bibinfo
  {volume} {4}},\ \bibinfo {pages} {2139} (\bibinfo {year} {1971})}\BibitemShut
  {NoStop}%
\bibitem [{\citenamefont {Abragam}\ and\ \citenamefont
  {Bleaney}(2012)}]{abragam2012electron}%
  \BibitemOpen
  \bibfield  {author} {\bibinfo {author} {\bibfnamefont {A.}~\bibnamefont
  {Abragam}}\ and\ \bibinfo {author} {\bibfnamefont {B.}~\bibnamefont
  {Bleaney}},\ }\href@noop {} {\emph {\bibinfo {title} {Electron paramagnetic
  resonance of transition ions}}}\ (\bibinfo  {publisher} {OUP Oxford},\
  \bibinfo {year} {2012})\BibitemShut {NoStop}%
\bibitem [{\citenamefont {Lines}(1963)}]{lines1963magnetic}%
  \BibitemOpen
  \bibfield  {author} {\bibinfo {author} {\bibfnamefont {M.~E.}\ \bibnamefont
  {Lines}},\ }\href@noop {} {\bibfield  {journal} {\bibinfo  {journal}
  {Physical Review}\ }\textbf {\bibinfo {volume} {131}},\ \bibinfo {pages}
  {546} (\bibinfo {year} {1963})}\BibitemShut {NoStop}%
\bibitem [{\citenamefont {Goff}\ \emph {et~al.}(1995)\citenamefont {Goff},
  \citenamefont {Tennant},\ and\ \citenamefont {Nagler}}]{goff1995exchange}%
  \BibitemOpen
  \bibfield  {author} {\bibinfo {author} {\bibfnamefont {J.~P.}\ \bibnamefont
  {Goff}}, \bibinfo {author} {\bibfnamefont {D.~A.}\ \bibnamefont {Tennant}}, \
  and\ \bibinfo {author} {\bibfnamefont {S.~E.}\ \bibnamefont {Nagler}},\
  }\href@noop {} {\bibfield  {journal} {\bibinfo  {journal} {Physical Review
  B}\ }\textbf {\bibinfo {volume} {52}},\ \bibinfo {pages} {15992} (\bibinfo
  {year} {1995})}\BibitemShut {NoStop}%
\bibitem [{\citenamefont {White}(2006)}]{white2006quantum}%
  \BibitemOpen
  \bibfield  {author} {\bibinfo {author} {\bibfnamefont {R.~M.}\ \bibnamefont
  {White}},\ }\href@noop {} {\emph {\bibinfo {title} {Quantum theory of
  magnetism}}},\ \bibinfo {edition} {3rd}\ ed.,\ edited by\ \bibinfo {editor}
  {\bibnamefont {Springer}}\ (\bibinfo  {publisher} {Springer},\ \bibinfo
  {year} {2006})\BibitemShut {NoStop}%
\bibitem [{\citenamefont {Ross}\ \emph {et~al.}(2011)\citenamefont {Ross},
  \citenamefont {Savary}, \citenamefont {Gaulin},\ and\ \citenamefont
  {Balents}}]{ross2011quantum}%
  \BibitemOpen
  \bibfield  {author} {\bibinfo {author} {\bibfnamefont {K.~A.}\ \bibnamefont
  {Ross}}, \bibinfo {author} {\bibfnamefont {L.}~\bibnamefont {Savary}},
  \bibinfo {author} {\bibfnamefont {B.~D.}\ \bibnamefont {Gaulin}}, \ and\
  \bibinfo {author} {\bibfnamefont {L.}~\bibnamefont {Balents}},\ }\href@noop
  {} {\bibfield  {journal} {\bibinfo  {journal} {Physical Review X}\ }\textbf
  {\bibinfo {volume} {1}},\ \bibinfo {pages} {021002} (\bibinfo {year}
  {2011})}\BibitemShut {NoStop}%
\bibitem [{\citenamefont {Guitteny}\ \emph {et~al.}(2013)\citenamefont
  {Guitteny}, \citenamefont {Petit}, \citenamefont {Lhotel}, \citenamefont
  {Robert}, \citenamefont {Bonville}, \citenamefont {Forget},\ and\
  \citenamefont {Mirebeau}}]{guitteny2013palmer}%
  \BibitemOpen
  \bibfield  {author} {\bibinfo {author} {\bibfnamefont {S.}~\bibnamefont
  {Guitteny}}, \bibinfo {author} {\bibfnamefont {S.}~\bibnamefont {Petit}},
  \bibinfo {author} {\bibfnamefont {E.}~\bibnamefont {Lhotel}}, \bibinfo
  {author} {\bibfnamefont {J.}~\bibnamefont {Robert}}, \bibinfo {author}
  {\bibfnamefont {P.}~\bibnamefont {Bonville}}, \bibinfo {author}
  {\bibfnamefont {A.}~\bibnamefont {Forget}}, \ and\ \bibinfo {author}
  {\bibfnamefont {I.}~\bibnamefont {Mirebeau}},\ }\href@noop {} {\bibfield
  {journal} {\bibinfo  {journal} {Physical Review B}\ }\textbf {\bibinfo
  {volume} {88}},\ \bibinfo {pages} {134408} (\bibinfo {year}
  {2013})}\BibitemShut {NoStop}%
\bibitem [{\citenamefont {Yan}\ \emph {et~al.}(2016)\citenamefont {Yan},
  \citenamefont {Benton}, \citenamefont {Jaubert},\ and\ \citenamefont
  {Shannon}}]{yan2016general}%
  \BibitemOpen
  \bibfield  {author} {\bibinfo {author} {\bibfnamefont {H.}~\bibnamefont
  {Yan}}, \bibinfo {author} {\bibfnamefont {O.}~\bibnamefont {Benton}},
  \bibinfo {author} {\bibfnamefont {L.~D.~C.}\ \bibnamefont {Jaubert}}, \ and\
  \bibinfo {author} {\bibfnamefont {N.}~\bibnamefont {Shannon}},\ }\href@noop
  {} {\bibfield  {journal} {\bibinfo  {journal} {arXiv preprint
  arXiv:1603.09466}\ } (\bibinfo {year} {2016})}\BibitemShut {NoStop}%
\bibitem [{\citenamefont {Champion}\ and\ \citenamefont
  {Holdsworth}(2004)}]{champion2004soft}%
  \BibitemOpen
  \bibfield  {author} {\bibinfo {author} {\bibfnamefont {J.~D.~M.}\
  \bibnamefont {Champion}}\ and\ \bibinfo {author} {\bibfnamefont {P.~C.~W.}\
  \bibnamefont {Holdsworth}},\ }\href@noop {} {\bibfield  {journal} {\bibinfo
  {journal} {Journal of Physics: Condensed Matter}\ }\textbf {\bibinfo {volume}
  {16}},\ \bibinfo {pages} {S665} (\bibinfo {year} {2004})}\BibitemShut
  {NoStop}%
\bibitem [{\citenamefont {Poole}\ \emph {et~al.}(2007)\citenamefont {Poole},
  \citenamefont {Wills},\ and\ \citenamefont
  {Lelievre-Berna}}]{poole2007magnetic}%
  \BibitemOpen
  \bibfield  {author} {\bibinfo {author} {\bibfnamefont {A.}~\bibnamefont
  {Poole}}, \bibinfo {author} {\bibfnamefont {A.~S.}\ \bibnamefont {Wills}}, \
  and\ \bibinfo {author} {\bibfnamefont {E.}~\bibnamefont {Lelievre-Berna}},\
  }\href@noop {} {\bibfield  {journal} {\bibinfo  {journal} {Journal of
  Physics: Condensed Matter}\ }\textbf {\bibinfo {volume} {19}},\ \bibinfo
  {pages} {452201} (\bibinfo {year} {2007})}\BibitemShut {NoStop}%
\bibitem [{\citenamefont {Petit}\ \emph {et~al.}(2014)\citenamefont {Petit},
  \citenamefont {Robert}, \citenamefont {Guitteny}, \citenamefont {Bonville},
  \citenamefont {Decorse}, \citenamefont {Ollivier}, \citenamefont {Mutka},
  \citenamefont {Gingras},\ and\ \citenamefont {Mirebeau}}]{petit2014order}%
  \BibitemOpen
  \bibfield  {author} {\bibinfo {author} {\bibfnamefont {S.}~\bibnamefont
  {Petit}}, \bibinfo {author} {\bibfnamefont {J.}~\bibnamefont {Robert}},
  \bibinfo {author} {\bibfnamefont {S.}~\bibnamefont {Guitteny}}, \bibinfo
  {author} {\bibfnamefont {P.}~\bibnamefont {Bonville}}, \bibinfo {author}
  {\bibfnamefont {C.}~\bibnamefont {Decorse}}, \bibinfo {author} {\bibfnamefont
  {J.}~\bibnamefont {Ollivier}}, \bibinfo {author} {\bibfnamefont
  {H.}~\bibnamefont {Mutka}}, \bibinfo {author} {\bibfnamefont {M.~J.}\
  \bibnamefont {Gingras}}, \ and\ \bibinfo {author} {\bibfnamefont
  {I.}~\bibnamefont {Mirebeau}},\ }\href@noop {} {\bibfield  {journal}
  {\bibinfo  {journal} {Physical Review B}\ }\textbf {\bibinfo {volume} {90}},\
  \bibinfo {pages} {060410} (\bibinfo {year} {2014})}\BibitemShut {NoStop}%
\bibitem [{\citenamefont {Gaudet}\ \emph {et~al.}(2016)\citenamefont {Gaudet},
  \citenamefont {Hallas}, \citenamefont {Maharaj}, \citenamefont {Buhariwalla},
  \citenamefont {Kermarrec}, \citenamefont {Butch}, \citenamefont {Munsie},
  \citenamefont {Dabkowska}, \citenamefont {Luke},\ and\ \citenamefont
  {Gaulin}}]{gaudet2016magnetic}%
  \BibitemOpen
  \bibfield  {author} {\bibinfo {author} {\bibfnamefont {J.}~\bibnamefont
  {Gaudet}}, \bibinfo {author} {\bibfnamefont {A.}~\bibnamefont {Hallas}},
  \bibinfo {author} {\bibfnamefont {D.}~\bibnamefont {Maharaj}}, \bibinfo
  {author} {\bibfnamefont {C.}~\bibnamefont {Buhariwalla}}, \bibinfo {author}
  {\bibfnamefont {E.}~\bibnamefont {Kermarrec}}, \bibinfo {author}
  {\bibfnamefont {N.}~\bibnamefont {Butch}}, \bibinfo {author} {\bibfnamefont
  {T.}~\bibnamefont {Munsie}}, \bibinfo {author} {\bibfnamefont
  {H.}~\bibnamefont {Dabkowska}}, \bibinfo {author} {\bibfnamefont
  {G.}~\bibnamefont {Luke}}, \ and\ \bibinfo {author} {\bibfnamefont
  {B.}~\bibnamefont {Gaulin}},\ }\href@noop {} {\bibfield  {journal} {\bibinfo
  {journal} {Physical Review B}\ }\textbf {\bibinfo {volume} {94}},\ \bibinfo
  {pages} {060407} (\bibinfo {year} {2016})}\BibitemShut {NoStop}%
\bibitem [{\citenamefont {Sarkar}\ \emph {et~al.}(2016)\citenamefont {Sarkar},
  \citenamefont {Krizan}, \citenamefont {Br{\"u}ckner}, \citenamefont
  {Andrade}, \citenamefont {Rachel}, \citenamefont {Vojta}, \citenamefont
  {Cava},\ and\ \citenamefont {Klauss}}]{sarkar2016unconventional}%
  \BibitemOpen
  \bibfield  {author} {\bibinfo {author} {\bibfnamefont {R.}~\bibnamefont
  {Sarkar}}, \bibinfo {author} {\bibfnamefont {J.~W.}\ \bibnamefont {Krizan}},
  \bibinfo {author} {\bibfnamefont {F.}~\bibnamefont {Br{\"u}ckner}}, \bibinfo
  {author} {\bibfnamefont {E.~C.}\ \bibnamefont {Andrade}}, \bibinfo {author}
  {\bibfnamefont {S.}~\bibnamefont {Rachel}}, \bibinfo {author} {\bibfnamefont
  {M.}~\bibnamefont {Vojta}}, \bibinfo {author} {\bibfnamefont {R.~J.}\
  \bibnamefont {Cava}}, \ and\ \bibinfo {author} {\bibfnamefont {H.-H.}\
  \bibnamefont {Klauss}},\ }\href@noop {} {\bibfield  {journal} {\bibinfo
  {journal} {arXiv preprint arXiv:1604.00814}\ } (\bibinfo {year}
  {2016})}\BibitemShut {NoStop}%
\bibitem [{\citenamefont {Ruff}\ \emph {et~al.}(2008)\citenamefont {Ruff},
  \citenamefont {Clancy}, \citenamefont {Bourque}, \citenamefont {White},
  \citenamefont {Ramazanoglu}, \citenamefont {Gardner}, \citenamefont {Qiu},
  \citenamefont {Copley}, \citenamefont {Johnson}, \citenamefont {Dabkowska}
  \emph {et~al.}}]{ruff2008spin}%
  \BibitemOpen
  \bibfield  {author} {\bibinfo {author} {\bibfnamefont {J.~P.~C.}\
  \bibnamefont {Ruff}}, \bibinfo {author} {\bibfnamefont {J.~P.}\ \bibnamefont
  {Clancy}}, \bibinfo {author} {\bibfnamefont {A.}~\bibnamefont {Bourque}},
  \bibinfo {author} {\bibfnamefont {M.~A.}\ \bibnamefont {White}}, \bibinfo
  {author} {\bibfnamefont {M.}~\bibnamefont {Ramazanoglu}}, \bibinfo {author}
  {\bibfnamefont {J.~S.}\ \bibnamefont {Gardner}}, \bibinfo {author}
  {\bibfnamefont {Y.}~\bibnamefont {Qiu}}, \bibinfo {author} {\bibfnamefont
  {J.~R.~D.}\ \bibnamefont {Copley}}, \bibinfo {author} {\bibfnamefont {M.~B.}\
  \bibnamefont {Johnson}}, \bibinfo {author} {\bibfnamefont {H.~A.}\
  \bibnamefont {Dabkowska}},  \emph {et~al.},\ }\href@noop {} {\bibfield
  {journal} {\bibinfo  {journal} {Physical review letters}\ }\textbf {\bibinfo
  {volume} {101}},\ \bibinfo {pages} {147205} (\bibinfo {year}
  {2008})}\BibitemShut {NoStop}%
\bibitem [{\citenamefont {Rodriguez}\ \emph {et~al.}(2008)\citenamefont
  {Rodriguez}, \citenamefont {Adler}, \citenamefont {Brand}, \citenamefont
  {Broholm}, \citenamefont {Cook}, \citenamefont {Brocker}, \citenamefont
  {Hammond}, \citenamefont {Huang}, \citenamefont {Hundertmark}, \citenamefont
  {Lynn} \emph {et~al.}}]{rodriguez2008macs}%
  \BibitemOpen
  \bibfield  {author} {\bibinfo {author} {\bibfnamefont {J.~A.}\ \bibnamefont
  {Rodriguez}}, \bibinfo {author} {\bibfnamefont {D.~M.}\ \bibnamefont
  {Adler}}, \bibinfo {author} {\bibfnamefont {P.~C.}\ \bibnamefont {Brand}},
  \bibinfo {author} {\bibfnamefont {C.}~\bibnamefont {Broholm}}, \bibinfo
  {author} {\bibfnamefont {J.~C.}\ \bibnamefont {Cook}}, \bibinfo {author}
  {\bibfnamefont {C.}~\bibnamefont {Brocker}}, \bibinfo {author} {\bibfnamefont
  {R.}~\bibnamefont {Hammond}}, \bibinfo {author} {\bibfnamefont
  {Z.}~\bibnamefont {Huang}}, \bibinfo {author} {\bibfnamefont
  {P.}~\bibnamefont {Hundertmark}}, \bibinfo {author} {\bibfnamefont {J.~W.}\
  \bibnamefont {Lynn}},  \emph {et~al.},\ }\href@noop {} {\bibfield  {journal}
  {\bibinfo  {journal} {Measurement Science and Technology}\ }\textbf {\bibinfo
  {volume} {19}},\ \bibinfo {pages} {034023} (\bibinfo {year}
  {2008})}\BibitemShut {NoStop}%
\bibitem [{\citenamefont {Granroth}\ \emph {et~al.}(2006)\citenamefont
  {Granroth}, \citenamefont {Vandergriff},\ and\ \citenamefont
  {Nagler}}]{granroth2006sequoia}%
  \BibitemOpen
  \bibfield  {author} {\bibinfo {author} {\bibfnamefont {G.~E.}\ \bibnamefont
  {Granroth}}, \bibinfo {author} {\bibfnamefont {D.~H.}\ \bibnamefont
  {Vandergriff}}, \ and\ \bibinfo {author} {\bibfnamefont {S.~E.}\ \bibnamefont
  {Nagler}},\ }\href@noop {} {\bibfield  {journal} {\bibinfo  {journal}
  {Physica B: Condensed Matter}\ }\textbf {\bibinfo {volume} {385}},\ \bibinfo
  {pages} {1104} (\bibinfo {year} {2006})}\BibitemShut {NoStop}%
\bibitem [{\citenamefont {Gaudet}\ \emph {et~al.}(2015)\citenamefont {Gaudet},
  \citenamefont {Maharaj}, \citenamefont {Sala}, \citenamefont {Kermarrec},
  \citenamefont {Ross}, \citenamefont {Dabkowska}, \citenamefont {Kolesnikov},
  \citenamefont {Granroth},\ and\ \citenamefont {Gaulin}}]{gaudet2015neutron}%
  \BibitemOpen
  \bibfield  {author} {\bibinfo {author} {\bibfnamefont {J.}~\bibnamefont
  {Gaudet}}, \bibinfo {author} {\bibfnamefont {D.~D.}\ \bibnamefont {Maharaj}},
  \bibinfo {author} {\bibfnamefont {G.}~\bibnamefont {Sala}}, \bibinfo {author}
  {\bibfnamefont {E.}~\bibnamefont {Kermarrec}}, \bibinfo {author}
  {\bibfnamefont {K.~A.}\ \bibnamefont {Ross}}, \bibinfo {author}
  {\bibfnamefont {H.~A.}\ \bibnamefont {Dabkowska}}, \bibinfo {author}
  {\bibfnamefont {A.~I.}\ \bibnamefont {Kolesnikov}}, \bibinfo {author}
  {\bibfnamefont {G.~E.}\ \bibnamefont {Granroth}}, \ and\ \bibinfo {author}
  {\bibfnamefont {B.~D.}\ \bibnamefont {Gaulin}},\ }\href@noop {} {\bibfield
  {journal} {\bibinfo  {journal} {Physical Review B}\ }\textbf {\bibinfo
  {volume} {92}},\ \bibinfo {pages} {134420} (\bibinfo {year}
  {2015})}\BibitemShut {NoStop}%
\bibitem [{\citenamefont {Babkevich}\ \emph {et~al.}(2015)\citenamefont
  {Babkevich}, \citenamefont {Finco}, \citenamefont {Jeong}, \citenamefont
  {Dalla~Piazza}, \citenamefont {Kovacevic}, \citenamefont {Klughertz},
  \citenamefont {Kr{\"a}mer}, \citenamefont {Kraemer}, \citenamefont {Adroja},
  \citenamefont {Goremychkin} \emph {et~al.}}]{babkevich2015neutron}%
  \BibitemOpen
  \bibfield  {author} {\bibinfo {author} {\bibfnamefont {P.}~\bibnamefont
  {Babkevich}}, \bibinfo {author} {\bibfnamefont {A.}~\bibnamefont {Finco}},
  \bibinfo {author} {\bibfnamefont {M.}~\bibnamefont {Jeong}}, \bibinfo
  {author} {\bibfnamefont {B.}~\bibnamefont {Dalla~Piazza}}, \bibinfo {author}
  {\bibfnamefont {I.}~\bibnamefont {Kovacevic}}, \bibinfo {author}
  {\bibfnamefont {G.}~\bibnamefont {Klughertz}}, \bibinfo {author}
  {\bibfnamefont {K.}~\bibnamefont {Kr{\"a}mer}}, \bibinfo {author}
  {\bibfnamefont {C.}~\bibnamefont {Kraemer}}, \bibinfo {author} {\bibfnamefont
  {D.~T.}\ \bibnamefont {Adroja}}, \bibinfo {author} {\bibfnamefont
  {E.}~\bibnamefont {Goremychkin}},  \emph {et~al.},\ }\href@noop {} {\bibfield
   {journal} {\bibinfo  {journal} {Physical Review B}\ }\textbf {\bibinfo
  {volume} {92}},\ \bibinfo {pages} {144422} (\bibinfo {year}
  {2015})}\BibitemShut {NoStop}%
\bibitem [{\citenamefont {Stevens}(1952)}]{stevens1952matrix}%
  \BibitemOpen
  \bibfield  {author} {\bibinfo {author} {\bibfnamefont {K.~W.~H.}\
  \bibnamefont {Stevens}},\ }\href@noop {} {\bibfield  {journal} {\bibinfo
  {journal} {Proceedings of the Physical Society. Section A}\ }\textbf
  {\bibinfo {volume} {65}},\ \bibinfo {pages} {209} (\bibinfo {year}
  {1952})}\BibitemShut {NoStop}%
\bibitem [{\citenamefont {Hutchings}(1964)}]{hutchings1964point}%
  \BibitemOpen
  \bibfield  {author} {\bibinfo {author} {\bibfnamefont {M.~T.}\ \bibnamefont
  {Hutchings}},\ }\href@noop {} {\bibfield  {journal} {\bibinfo  {journal}
  {Solid state physics}\ }\textbf {\bibinfo {volume} {16}},\ \bibinfo {pages}
  {227} (\bibinfo {year} {1964})}\BibitemShut {NoStop}%
\bibitem [{\citenamefont {Rotter}\ \emph {et~al.}()\citenamefont {Rotter},
  \citenamefont {Manh~Le}, \citenamefont {Keller}, \citenamefont {Pascut},
  \citenamefont {Hoffmann}, \citenamefont {Doerr}, \citenamefont {Schedler},
  \citenamefont {Fabi}, \citenamefont {Rotter}, \citenamefont {Banks},\ and\
  \citenamefont {Kluver}}]{mcphase_manual}%
  \BibitemOpen
  \bibfield  {author} {\bibinfo {author} {\bibfnamefont {M.}~\bibnamefont
  {Rotter}}, \bibinfo {author} {\bibfnamefont {D.}~\bibnamefont {Manh~Le}},
  \bibinfo {author} {\bibfnamefont {J.}~\bibnamefont {Keller}}, \bibinfo
  {author} {\bibfnamefont {L.~G.}\ \bibnamefont {Pascut}}, \bibinfo {author}
  {\bibfnamefont {T.}~\bibnamefont {Hoffmann}}, \bibinfo {author}
  {\bibfnamefont {M.}~\bibnamefont {Doerr}}, \bibinfo {author} {\bibfnamefont
  {R.}~\bibnamefont {Schedler}}, \bibinfo {author} {\bibfnamefont
  {P.}~\bibnamefont {Fabi}}, \bibinfo {author} {\bibfnamefont {S.}~\bibnamefont
  {Rotter}}, \bibinfo {author} {\bibfnamefont {M.}~\bibnamefont {Banks}}, \
  and\ \bibinfo {author} {\bibfnamefont {N.}~\bibnamefont {Kluver}},\ }\href
  {http://www2.cpfs.mpg.de/~rotter/homepage_mcphase/manual/manual.html}
  {\enquote {\bibinfo {title} {Mcphase users manual: Symmetry considerations
  for crystal field parameters},}\ }\BibitemShut {NoStop}%
\bibitem [{\citenamefont {Kanamori}(1957)}]{kanamori1957theory}%
  \BibitemOpen
  \bibfield  {author} {\bibinfo {author} {\bibfnamefont {J.}~\bibnamefont
  {Kanamori}},\ }\href@noop {} {\bibfield  {journal} {\bibinfo  {journal}
  {Progress of Theoretical Physics}\ }\textbf {\bibinfo {volume} {17}},\
  \bibinfo {pages} {177} (\bibinfo {year} {1957})}\BibitemShut {NoStop}%
\bibitem [{\citenamefont {Squires}(2012)}]{squires2012introduction}%
  \BibitemOpen
  \bibfield  {author} {\bibinfo {author} {\bibfnamefont {G.~L.}\ \bibnamefont
  {Squires}},\ }\href@noop {} {\emph {\bibinfo {title} {Introduction to the
  theory of thermal neutron scattering}}}\ (\bibinfo  {publisher} {Cambridge
  university press},\ \bibinfo {year} {2012})\BibitemShut {NoStop}%
\bibitem [{\citenamefont {Jensen}\ and\ \citenamefont
  {Mackintosh}(1991)}]{jensen1991rare}%
  \BibitemOpen
  \bibfield  {author} {\bibinfo {author} {\bibfnamefont {J.}~\bibnamefont
  {Jensen}}\ and\ \bibinfo {author} {\bibfnamefont {A.~R.}\ \bibnamefont
  {Mackintosh}},\ }\href@noop {} {\emph {\bibinfo {title} {Rare earth
  magnetism}}}\ (\bibinfo  {publisher} {Clarendon Oxford},\ \bibinfo {year}
  {1991})\BibitemShut {NoStop}%
\bibitem [{\citenamefont {Oliveira}\ \emph {et~al.}(2004)\citenamefont
  {Oliveira}, \citenamefont {Guedes}, \citenamefont {Ayala}, \citenamefont
  {Gesland}, \citenamefont {Ellena}, \citenamefont {Moreira},\ and\
  \citenamefont {Grimsditch}}]{oliveira2004crystal}%
  \BibitemOpen
  \bibfield  {author} {\bibinfo {author} {\bibfnamefont {E.~A.}\ \bibnamefont
  {Oliveira}}, \bibinfo {author} {\bibfnamefont {I.}~\bibnamefont {Guedes}},
  \bibinfo {author} {\bibfnamefont {A.~P.}\ \bibnamefont {Ayala}}, \bibinfo
  {author} {\bibfnamefont {J.~Y.}\ \bibnamefont {Gesland}}, \bibinfo {author}
  {\bibfnamefont {J.}~\bibnamefont {Ellena}}, \bibinfo {author} {\bibfnamefont
  {R.~L.}\ \bibnamefont {Moreira}}, \ and\ \bibinfo {author} {\bibfnamefont
  {M.}~\bibnamefont {Grimsditch}},\ }\href@noop {} {\bibfield  {journal}
  {\bibinfo  {journal} {Journal of Solid State Chemistry}\ }\textbf {\bibinfo
  {volume} {177}},\ \bibinfo {pages} {2943} (\bibinfo {year}
  {2004})}\BibitemShut {NoStop}%
\bibitem [{\citenamefont {Kimura}\ \emph {et~al.}(2014)\citenamefont {Kimura},
  \citenamefont {Nakatsuji},\ and\ \citenamefont
  {Kimura}}]{kimura2014experimental}%
  \BibitemOpen
  \bibfield  {author} {\bibinfo {author} {\bibfnamefont {K.}~\bibnamefont
  {Kimura}}, \bibinfo {author} {\bibfnamefont {S.}~\bibnamefont {Nakatsuji}}, \
  and\ \bibinfo {author} {\bibfnamefont {T.}~\bibnamefont {Kimura}},\
  }\href@noop {} {\bibfield  {journal} {\bibinfo  {journal} {Physical Review
  B}\ }\textbf {\bibinfo {volume} {90}},\ \bibinfo {pages} {060414} (\bibinfo
  {year} {2014})}\BibitemShut {NoStop}%
\bibitem [{\citenamefont {Baroudi}\ \emph {et~al.}(2015)\citenamefont
  {Baroudi}, \citenamefont {Gaulin}, \citenamefont {Lapidus}, \citenamefont
  {Gaudet},\ and\ \citenamefont {Cava}}]{baroudi2015symmetry}%
  \BibitemOpen
  \bibfield  {author} {\bibinfo {author} {\bibfnamefont {K.}~\bibnamefont
  {Baroudi}}, \bibinfo {author} {\bibfnamefont {B.~D.}\ \bibnamefont {Gaulin}},
  \bibinfo {author} {\bibfnamefont {S.~H.}\ \bibnamefont {Lapidus}}, \bibinfo
  {author} {\bibfnamefont {J.}~\bibnamefont {Gaudet}}, \ and\ \bibinfo {author}
  {\bibfnamefont {R.~J.}\ \bibnamefont {Cava}},\ }\href@noop {} {\bibfield
  {journal} {\bibinfo  {journal} {Physical Review B}\ }\textbf {\bibinfo
  {volume} {92}},\ \bibinfo {pages} {024110} (\bibinfo {year}
  {2015})}\BibitemShut {NoStop}%
\bibitem [{Note1()}]{Note1}%
  \BibitemOpen
  \bibinfo {note} {The $E \sim 360$ meV mode was measured only in NCCF due to
  time limitations. Given the overall similarity if the other single-ion levels
  in the two materials, this mode is assumed to lie in same energy range in
  NSCF.}\BibitemShut {Stop}%
\bibitem [{\citenamefont {Campbell}\ \emph {et~al.}(2006)\citenamefont
  {Campbell}, \citenamefont {Stokes}, \citenamefont {Tanner},\ and\
  \citenamefont {Hatch}}]{campbell2006isodisplace}%
  \BibitemOpen
  \bibfield  {author} {\bibinfo {author} {\bibfnamefont {B.~J.}\ \bibnamefont
  {Campbell}}, \bibinfo {author} {\bibfnamefont {H.~T.}\ \bibnamefont
  {Stokes}}, \bibinfo {author} {\bibfnamefont {D.~E.}\ \bibnamefont {Tanner}},
  \ and\ \bibinfo {author} {\bibfnamefont {D.~M.}\ \bibnamefont {Hatch}},\
  }\href@noop {} {\bibfield  {journal} {\bibinfo  {journal} {Journal of Applied
  Crystallography}\ }\textbf {\bibinfo {volume} {39}},\ \bibinfo {pages} {607}
  (\bibinfo {year} {2006})}\BibitemShut {NoStop}%
\bibitem [{\citenamefont {Shannon}(1976)}]{shannon1976revised}%
  \BibitemOpen
  \bibfield  {author} {\bibinfo {author} {\bibfnamefont {R.~D.}\ \bibnamefont
  {Shannon}},\ }\href@noop {} {\bibfield  {journal} {\bibinfo  {journal} {Acta
  Crystallographica Section A: Crystal Physics, Diffraction, Theoretical and
  General Crystallography}\ }\textbf {\bibinfo {volume} {32}},\ \bibinfo
  {pages} {751} (\bibinfo {year} {1976})}\BibitemShut {NoStop}%
\bibitem [{\citenamefont {Tomiyasu}\ \emph {et~al.}(2011)\citenamefont
  {Tomiyasu}, \citenamefont {Crawford}, \citenamefont {Adroja}, \citenamefont
  {Manuel}, \citenamefont {Tominaga}, \citenamefont {Hara}, \citenamefont
  {Sato}, \citenamefont {Watanabe}, \citenamefont {Ikeda}, \citenamefont {Lynn}
  \emph {et~al.}}]{tomiyasu2011molecular}%
  \BibitemOpen
  \bibfield  {author} {\bibinfo {author} {\bibfnamefont {K.}~\bibnamefont
  {Tomiyasu}}, \bibinfo {author} {\bibfnamefont {M.~K.}\ \bibnamefont
  {Crawford}}, \bibinfo {author} {\bibfnamefont {D.~T.}\ \bibnamefont
  {Adroja}}, \bibinfo {author} {\bibfnamefont {P.}~\bibnamefont {Manuel}},
  \bibinfo {author} {\bibfnamefont {A.}~\bibnamefont {Tominaga}}, \bibinfo
  {author} {\bibfnamefont {S.}~\bibnamefont {Hara}}, \bibinfo {author}
  {\bibfnamefont {H.}~\bibnamefont {Sato}}, \bibinfo {author} {\bibfnamefont
  {T.}~\bibnamefont {Watanabe}}, \bibinfo {author} {\bibfnamefont {S.~I.}\
  \bibnamefont {Ikeda}}, \bibinfo {author} {\bibfnamefont {J.~W.}\ \bibnamefont
  {Lynn}},  \emph {et~al.},\ }\href@noop {} {\bibfield  {journal} {\bibinfo
  {journal} {Physical Review B}\ }\textbf {\bibinfo {volume} {84}},\ \bibinfo
  {pages} {054405} (\bibinfo {year} {2011})}\BibitemShut {NoStop}%
\bibitem [{\citenamefont {Arnold}\ \emph {et~al.}(2014)\citenamefont {Arnold},
  \citenamefont {Bilheux}, \citenamefont {Borreguero}, \citenamefont {Buts},
  \citenamefont {Campbell}, \citenamefont {Chapon}, \citenamefont {Doucet},
  \citenamefont {Draper}, \citenamefont {Leal}, \citenamefont {Gigg} \emph
  {et~al.}}]{arnold2014mantid}%
  \BibitemOpen
  \bibfield  {author} {\bibinfo {author} {\bibfnamefont {O.}~\bibnamefont
  {Arnold}}, \bibinfo {author} {\bibfnamefont {J.-C.}\ \bibnamefont {Bilheux}},
  \bibinfo {author} {\bibfnamefont {J.}~\bibnamefont {Borreguero}}, \bibinfo
  {author} {\bibfnamefont {A.}~\bibnamefont {Buts}}, \bibinfo {author}
  {\bibfnamefont {S.~I.}\ \bibnamefont {Campbell}}, \bibinfo {author}
  {\bibfnamefont {L.}~\bibnamefont {Chapon}}, \bibinfo {author} {\bibfnamefont
  {M.}~\bibnamefont {Doucet}}, \bibinfo {author} {\bibfnamefont
  {N.}~\bibnamefont {Draper}}, \bibinfo {author} {\bibfnamefont {R.~F.}\
  \bibnamefont {Leal}}, \bibinfo {author} {\bibfnamefont {M.~A.}\ \bibnamefont
  {Gigg}},  \emph {et~al.},\ }\href@noop {} {\bibfield  {journal} {\bibinfo
  {journal} {Nuclear Instruments and Methods in Physics Research Section A:
  Accelerators, Spectrometers, Detectors and Associated Equipment}\ }\textbf
  {\bibinfo {volume} {764}},\ \bibinfo {pages} {156} (\bibinfo {year}
  {2014})}\BibitemShut {NoStop}%
\bibitem [{\citenamefont {Azuah}\ \emph {et~al.}(2009)\citenamefont {Azuah},
  \citenamefont {Kneller}, \citenamefont {Qiu}, \citenamefont
  {Tregenna-Piggott}, \citenamefont {Brown}, \citenamefont {Copley},\ and\
  \citenamefont {Dimeo}}]{azuah2009dave}%
  \BibitemOpen
  \bibfield  {author} {\bibinfo {author} {\bibfnamefont {R.~T.}\ \bibnamefont
  {Azuah}}, \bibinfo {author} {\bibfnamefont {L.~R.}\ \bibnamefont {Kneller}},
  \bibinfo {author} {\bibfnamefont {Y.}~\bibnamefont {Qiu}}, \bibinfo {author}
  {\bibfnamefont {P.~L.}\ \bibnamefont {Tregenna-Piggott}}, \bibinfo {author}
  {\bibfnamefont {C.~M.}\ \bibnamefont {Brown}}, \bibinfo {author}
  {\bibfnamefont {J.~R.}\ \bibnamefont {Copley}}, \ and\ \bibinfo {author}
  {\bibfnamefont {R.~M.}\ \bibnamefont {Dimeo}},\ }\href@noop {} {\bibfield
  {journal} {\bibinfo  {journal} {Journal of Research of the National Institute
  of Standards and Technology}\ }\textbf {\bibinfo {volume} {114}},\ \bibinfo
  {pages} {341} (\bibinfo {year} {2009})}\BibitemShut {NoStop}%
\bibitem [{\citenamefont {Lovesey}(1984)}]{lovesey1984theory}%
  \BibitemOpen
  \bibfield  {author} {\bibinfo {author} {\bibfnamefont {S.~W.}\ \bibnamefont
  {Lovesey}},\ }\href@noop {} {\emph {\bibinfo {title} {Theory of neutron
  scattering from condensed matter}}},\ Vol.~\bibinfo {volume} {2}\ (\bibinfo
  {address} {United Kingdom},\ \bibinfo {year} {1984})\BibitemShut {NoStop}%
\end{thebibliography}

\end{document}